\documentclass[fleqn,usenatbib]{mnras}

\usepackage{newtxtext,newtxmath}

\usepackage[T1]{fontenc}

\DeclareRobustCommand{\VAN}[3]{#2}
\let\VANthebibliography\thebibliography
\def\thebibliography{\DeclareRobustCommand{\VAN}[3]{##3}\VANthebibliography}

\usepackage{graphicx}	
\usepackage{amsmath}	
\usepackage{lscape} 
\usepackage{subfigure}
\usepackage{caption}
\usepackage{multirow}
\usepackage{float}
\usepackage{adjustbox}
\usepackage{placeins}
\usepackage{lineno}

\makeatletter
\newcommand{\labtext}[2]{%
  \@bsphack
  \csname phantomsection\endcsname 
  \def\@currentlabel{#1}{\label{#2}}%
  \@esphack
}
\makeatother

\title[Optical polarization of $\gamma$-ray blazars]{Variability and evolution of the optical polarization of a sample of gamma-ray blazars}

\author[J. Otero-Santos et al.]{
J. Otero-Santos,$^{1,2}$\thanks{E-mail: joteros@iac.es (JOS), jap@iac.es (JAP), jbecerra@iac.es (JBG)}\thanks{Recently moved to Instituto de Astrofísica de Andalucía, CSIC, E-18080, Granada, Spain}
J. A. Acosta-Pulido,$^{1,2~\textcolor{blue}{\star}}$
J. Becerra Gonz\'alez,$^{1,2~\textcolor{blue}{\star}}$
C. M. Raiteri,$^{3}$
M. I. Carnerero,$^{3}$
\newauthor
N. Castro Segura,$^{2,4}$
O. González-Martín$^{5}$ and
A. Luashvili$^{6}$ 
\\
$^{1}$Instituto de Astrof\'isica de Canarias (IAC), E-38200 La Laguna, Tenerife, Spain\\
$^{2}$Universidad de La Laguna (ULL), Departamento de Astrof\'isica, E-38206 La Laguna, Tenerife, Spain\\
$^{3}$INAF-Osservatorio Astrofisico di Torino, via Osservatorio 20, 10025 Pino Torinese, Italy\\
$^{4}$School of Physics \& Astronomy, University of Southampton, Highfield, Southampton SO17 1BJ, UK\\
$^{5}$Instituto de Radioastronomía y Astrofísica (IRyA-UNAM), 3-72 (Xangari), 8701, Morelia, Mexico\\
$^{6}$Laboratoire Univers et Théories, Observatoire de Paris, Unversité PSL, CNRS, Université de Paris, 92190 Meudon, France\\
} 

\date{Accepted XXX. Received YYY; in original form ZZZ}

\pubyear{2023}

\begin{document}
\label{firstpage}
\pagerange{\pageref{firstpage}--\pageref{lastpage}}
\maketitle

\begin{abstract}
We present a polarization variability analysis of a sample of 26 $\gamma$-ray blazars monitored by the Steward Observatory between 2008 and 2018 in the optical band. We investigate the properties and long-term variability of their optical polarization, searching for differences between blazar types. We observe that BL Lac objects are typically less polarized and less variable than flat spectrum radio quasars (FSRQs). Moreover, BL Lacs display a distribution of their polarization angle typically oriented in a preferential direction, contrary to the rather random distribution of FSRQs. For the latter blazar type, as well as those sources showing a bright stellar emission, we take into account the depolarizing effect introduced by the broad line region and the host galaxy on the measured polarization degree. {In this sample} we also observe that BL Lacs present an uncorrelated evolution of the flux and the polarization. Contrary, FSRQs show a correlation before the depolarization correction, that is lost however after considering this effect. In addition, we study the behaviour of the polarization angle, searching for angle rotations in its long-term evolution. We derive that {the FSRQs studied here} show rotations more frequently than BL Lac objects by a factor $\sim$1.5. During these periods we also observe a systematic decrease of the polarization fraction, as well as a marginal flux increase, not significant however to connect rotations with optical flares. We interpret these results within the extended shock-in-jet scenario, able to explain the overall features observed here for the polarization of the blazar sample.

\end{abstract}




\section{Introduction}\label{sec:introduction}
Active galactic nuclei (AGN) are one of the most powerful sources in the entire Universe, sometimes with the ability of developing powerful relativistic jets acting as natural particle accelerators. Among the different types, blazars are extremely boosted due to relativistic effects, owing to their orientation being closely aligned with the {line of sight} of the Earth. They can emit from radio to very-high-energy $\gamma$-rays, displaying a characteristic double-bump shape in their spectral energy distribution \citep[SED, see e.g.][]{abdo2010}. The low energy bump, {with its peak} typically found {at} infrared (IR) to X-ray frequencies, is explained with non-thermal synchrotron emission of relativistic electrons from the jet \citep{ghisellini2010}. The high energy bump found at $\gamma$-ray energies on the other hand is {often} interpreted as an inverse Compton (IC) scattering of low energy photons with the same population of relativistic electrons responsible for the synchrotron emission \citep{maraschi1992}. Blazars are often subdivided in two classes, depending on their optical spectrum \citep{urry1995}: BL Lacertae (BL Lac) objects and flat spectrum radio quasars (FSRQs). The former present an optical spectrum with no features or weak features with an equivalent width $|\text{EW}|<5$~\AA \ in the rest frame, while in the optical spectrum of the latter type, narrow and broad lines with $|\text{EW}|>5$~\AA \ are visible.

The low energy emission of AGN, from radio to optical wavelengths, is characterised by a high polarization degree. This polarized emission is related to the non-thermal synchrotron emission and the magnetic field of the relativistically boosted jet. Among the large variety of AGN types, blazars show an extraordinarily high optical polarization w.r.t. other classes, as reported by \cite{angelakis2016}, with values that can reach a fraction of $\sim$50\% \citep{smith2017}. This polarization is affected by the characteristic intense variability displayed by blazars, exhibiting for instance large changes of the polarization fraction or the orientation of the polarization angle. Owing to the origin of this polarized emission, this variability is linked to the magnetic field of the jet and its variations. As a consequence, variability studies of the polarization of blazars are one of the most accessible ways of obtaining information of the magnetic field in these objects and in the extreme Universe, the role it plays in the particle acceleration, the interaction with its environment, and its relation with the overall emission detected from blazars. The polarized emission has been observed in the past with many different behaviours w.r.t. the total optical emission. For instance, \cite{raiteri2012} reported a correlated evolution of the optical total and polarized flux for the FSRQ B2~1633+38. On the other hand, anticorrelations between the total and polarized emission of blazars have also been reported, e.g. \cite{fraija2017} for Mkn~421. In addition, the lack of relation between these two quantities has also been observed in some cases \citep[e.g. for PKS~1424+240 as reported by][]{covino2015}.

{In this regard, the RoboPol programme has been leading the effort on studying the polarized emission of astrophysical objects and in particular of AGN and blazars. These studies have evaluated different characteristics of the polarized emission of these objects and their variability in a systematic way. For instance, \cite{angelakis2016} have studied the relation between the frequency of the synchrotron peak and the radio loudness of several AGN. Properties such as variability and distribution of the polarization in AGN and blazars have also been characterized by e.g. \cite{pavlidou2014,hovatta2016}. Numerous efforts have also been dedicated to understand the development of rotations of the polarization angle \citep[see][]{blinov2015,blinov2016a,blinov2016b,blinov2018}, as well as their connection with the $\gamma$-ray activity of blazars and/or whether they have a stochastic origin \citep{blinov2015,blinov2018}.}

Studying the polarized emission and variability of $\gamma$-ray blazars is also key to understand the connection of the magnetic field with the mechanisms responsible for the broadband emission and the particle acceleration in the jet. However, the fact that each blazar seems to behave in its own unique way makes the interpretation within a general framework rather difficult, leading to a large variety of models developed for understanding their behaviour. For example, models based on chaotic magnetic fields have been proposed by \cite{laing1980}. Moreover, helical jets with helical magnetic fields like the model developed by \cite{raiteri2013} have been commonly used to explain the evolution of the polarization of blazars \citep[see e.g.][]{gupta2017}. \cite{marscher2010} suggest a model based on knots moving under the influence of a toroidal magnetic field following a spiral path. Another possibility involves magnetic reconnection of oppositely directed field lines, releasing magnetic energy and accelerating particles \citep{zhang2018}. Alternatively, models based on turbulent emitting cells in the jet have also been widely used scenario for interpreting the polarization variability \cite[see][]{marscher2014}.


Here we study the behaviour, variability and evolution of the polarization and its characteristics for a sample of $\gamma$-ray bright blazars. These blazars have been regularly monitored by the Steward Observatory for approximately 10 years, providing an excellent spectropolarimetric database for a long-term analysis of the properties and variability of their polarization. {This database has been explored in the past for individual sources. However, a systematic study of a large sample of sources observed by this monitoring programme is still lacking. Here we perform an extensive evaluation of the polarized emission of our blazar sample owing to the remarkable time coverage and sampling, only comparable to the RoboPol programme, that has observed a larger number of sources with however a shorter coverage of $\sim$3~years \citep[see e.g.][]{blinov2016a,blinov2016b,blinov2018}. This allows a thorough comparison with the results derived by RoboPol. In addition, taking advantage of the spectropolarimetry performed by the Steward Observatory, a comparison of the behaviour of the whole spectral dataset studied in \cite{otero-santos2022} (hereafter OS22) and the polarized spectra is also conducted.} The paper is structured as follows: in Section \ref{sec:observations} we detail the observations and data reduction used in this analysis; the methodology followed for the data analysis is explained in Section \ref{sec:methodology}; and the main results and discussion are detailed in Sections \ref{sec:results} and \ref{sec:discussion}, respectively. Finally, a brief summary of the main conclusions is also included in Section \ref{sec:conclusions}.

\section{Observations and data reduction}\label{sec:observations}
In this work we have included the polarimetric data of 26 $\gamma$-ray bright blazars monitored by the Steward Observatory and previously analysed in OS22. These sources were observed for roughly 10 years, between 2008 and 2018, with spectropolarimetric observations in support of the \textit{Fermi}-LAT telescope. The Steward Observatory provides the optical total and polarized spectra between 4000 and 7550~\AA, with resolutions ranging from 16 to 24~\AA \ and a dispersion of approximately 4~\AA/pixel, depending on the slit width. In addition, the average polarization values, including the Stokes parameters, the polarization degree and the orientation of the electric vector position angle (EVPA) in the range of 5000-7000~\AA \ are also provided. The blazar sample analysed here is presented in Table \ref{tab:blazar_sample}, separated {by} their blazar type as BL Lacs, FSRQs, and BL Lacs dominated by the emission of the stellar population according to OS22. We have also included their classification according to the position of the synchrotron peak of the SED as low- (LSPs, $\nu_{\rm sync}$~<~10$^{14}$\,Hz), intermediate- (ISPs, 10$^{14}$\,Hz~<~$\nu_{\rm sync}$~<~10$^{15}$\,Hz) or high-synchrotron peaked (HSPs, $\nu_{\rm sync}$~>~10$^{15}$\,Hz) blazars, as defined in \cite{ajello2020}. 
Following the procedure from OS22, we have eliminated the telluric absorption features, performed the rest frame conversion, corrected small shifts in $\lambda$ and accounted for the Galactic extinction using the model from \cite{fitzpatrick1999}. More details are given in Section 2 of OS22.

\begin{table}
\begin{center}
\caption{Sample of blazars from the Steward Observatory monitoring programme included in this work.}
\begin{tabular}{ccccc} \hline \hline
Source & Sp. Type & $z$ & SED Type$^{a}$ & $\nu_{sync}$$^{a}$  \\ \hline \hline
H 1426+428 & \begin{tabular}[c]{@{}c@{}}Galaxy\\dominated\end{tabular} & 0.129 & HSP & 1.0$\cdot$10$^{18}$ \\ \hline
Mkn 501 & \begin{tabular}[c]{@{}c@{}}Galaxy\\dominated\end{tabular} & 0.033 & HSP & 2.8$\cdot$10$^{15}$ \\ \hline
1ES 2344+514 & \begin{tabular}[c]{@{}c@{}}Galaxy\\dominated\end{tabular} &  0.044 & HSP & 1.6$\cdot$10$^{16}$ \\ \hline \hline
3C 66A & BL Lac & 0.444$^{b}$ & ISP & 7.1$\cdot$10$^{14}$ \\ \hline
AO 0235+164 & BL Lac & 0.940 & LSP & 2.5$\cdot$10$^{14}$ \\ \hline
S5 0716+714 & BL Lac & 0.30$^{c}$ & ISP & 1.5$\cdot$10$^{14}$ \\ \hline
PKS 0735+178 & BL Lac & 0.424 & LSP & 2.7$\cdot$10$^{13}$ \\ \hline
OJ 287 & BL Lac & 0.306 & LSP & 1.8$\cdot$10$^{13}$  \\ \hline
Mkn 421 & BL Lac & 0.031 & HSP & 1.7$\cdot$10$^{16}$ \\ \hline
W Comae & BL Lac & 0.103 & ISP & 4.5$\cdot$10$^{14}$  \\ \hline
H 1219+305 & BL Lac & 0.184 & HSP & 1.9$\cdot$10$^{16}$ \\ \hline
1ES 1959+650 & BL Lac & 0.047 & HSP & 9.0$\cdot$10$^{15}$ \\ \hline
PKS 2155-304$^{\dagger}$ & BL Lac & 0.116 & HSP & 5.7$\cdot$10$^{15}$ \\ \hline
BL Lacertae & BL Lac & 0.069 & LSP & 3.9$\cdot$10$^{13}$ \\ \hline \hline
PKS 0420-014 & FSRQ &  0.916 & LSP & 5.9$\cdot$10$^{12}$ \\ \hline
PKS 0736+017$^{\dagger}$ & FSRQ &  0.189 & LSP & 1.5$\cdot$10$^{13}$ \\ \hline
OJ 248 & FSRQ & 0.941 & LSP & 3.4$\cdot$10$^{12}$ \\ \hline
Ton 599 & FSRQ & 0.725 & LSP & 1.2$\cdot$10$^{13}$ \\ \hline
PKS 1222+216 & FSRQ &  0.434 & LSP & 2.9$\cdot$10$^{13}$ \\ \hline
3C 273 & FSRQ & 0.158 & LSP & 7.0$\cdot$10$^{13}$ \\ \hline
3C 279 & FSRQ & 0.536 & LSP & 5.2$\cdot$10$^{12}$ \\ \hline
PKS 1510-089 & FSRQ &  0.360 & LSP & 1.1$\cdot$10$^{13}$ \\ \hline
B2 1633+38 & FSRQ & 1.814 & LSP & 3.0$\cdot$10$^{12}$ \\ \hline
3C 345 & FSRQ & 0.593 & LSP & 1.0$\cdot$10$^{13}$ \\ \hline
CTA 102 & FSRQ & 1.032 & LSP & 3.0$\cdot$10$^{12}$ \\ \hline
3C 454.3 & FSRQ & 0.859 & LSP & 1.3$\cdot$10$^{13}$ \\ \hline \hline
\end{tabular}
\label{tab:blazar_sample}
\end{center}
\footnotesize{$^{a}$Extracted from the 4LAC-DR2 catalogue of the \textit{Fermi}-LAT satellite \citep{ajello2020}.

$^{b}$Redshift value still under debate. Lower limit of $z \geqslant 0.33$ determined by \cite{torres-zafra2018}. 

$^{c}$Redshift still uncertain. Upper limit of $z < 0.322$ estimated by \cite{danforth2013}.

$^{\dagger}$Sources that have not been observed by the RoboPol programme between 2013 and 2017.}
\end{table}


In addition to the Galactic extinction, the interstellar medium can also have a small contribution to the observed polarization degree. We have evaluated this contribution for the sources with the highest expected interstellar polarization, that depends on the extinction value $A_{\lambda}$, i.e. 1ES~1959+650 ($A_{\lambda}=0.474$), BL~Lacertae ($A_{\lambda}=0.901$) and 1ES~2344+514 ($A_{\lambda}=0.580$). We have used the approximation from \cite{serkowski1975}, where the interstellar contribution can be expressed as $P_{V}\sim4.5 E_{B-V}$. In this expression, $E_{B-V}$ represents the reddening of the sources, $E_{B-V}=A_{\lambda}/R_{V}$. We adopted a value of $R_{V}=3.1$ and retrieved the values of $A_{\lambda}$ for these sources from \cite{schlafly2011}. With these considerations, the expected interstellar medium polarization is 0.67\%, 1.3\% and 0.84\%, respectively. To crosscheck this result, we have made use of the catalogue from \cite{heiles2000} to estimate the contribution of the interstellar medium from the reddening of a known star located close to the position of the source. This alternative approach leads to contributions of $\sim$0.78\%, $\sim$0.65\% and $\sim$0.19\%, respectively. Therefore, we assume that the interstellar polarization contribution is negligible for the present work.

{The ambiguity on the EVPA value was taken into account with the approach used in other studies \citep[e.g.][]{kiehlmann2016}. We consider that consecutive measurements of the polarization angle correspond to those that minimize the difference between them by considering the criterion}
\begin{equation}
\Delta \theta_{i} =  |\theta_{i+1}-\theta_{i}| < 90^{\circ}.
\label{eq:EVPA_ambiguity}
\end{equation}
{Following this equation, $\pm$n$\cdot$180$^{\circ}$ is added to $\theta_{i+1}$ if $\Delta \theta_{i}$>90$^{\circ}$. We note that other authors \citep[for instance,][]{carnerero2017} include the uncertainties in the determination of the EVPA when performing this correction,}
\begin{equation}
\Delta \theta_{i} =  |\theta_{i+1}-\theta_{i}| - \sqrt{\sigma(\theta_{i+1})^2+\sigma(\theta_{i})^2} < 90^{\circ}.
\label{eq:EVPA_ambiguity_2}
\end{equation}
{This can lead to a correction between two measurements that would not be performed with the criterion of Equation (\ref{eq:EVPA_ambiguity}). For the present work, we have adopted the first criterion presented here. Nevertheless, we have checked both criteria, with no different results in this case. Moreover, this correction is also dependant on the initial choice of EVPA interval, i.e. [-90$^{\circ}$, 90$^{\circ}$] or [0$^{\circ}$, 180$^{\circ}$]. Here we use the latter, that corresponds to the interval used by the Steward Observatory data.}
We stress that our data are affected by seasonal observational gaps. Since blazars are characterised by a strong variability {on} many different time scales \citep[see e.g.][]{kiehlmann2021}, we only apply this condition to measurements taken with a temporal separation <200~days.

{The blazars monitored by the Steward Observatory have been selected based on their interest as emitting \textit{Fermi}-LAT $\gamma$-ray sources with extreme variability. This could lead to the introduction of a bias towards $\gamma$-ray bright sources in the present case. This contrasts with the criterion used by, for instance, the RoboPol programme, that consists on an unbiased sample of $\gamma$-ray loud blazars \citep[see][]{pavlidou2014,blinov2015}. Nevertheless, we also note that all sources included here except PKS~2155-304 and PKS~0736+017 (as shown in Table \ref{tab:blazar_sample}), are coincident with sources observed by RoboPol.}

\section{Methodology}\label{sec:methodology}
The aim of this work is to study the properties, behaviour and evolution of the polarization for the blazar sample considered here. For this, we have evaluated the variability and the overall behaviour of the polarization degree and angle (Section \ref{sec:3.1}). Moreover, as the host galaxy and broad line region (BLR) from sources with a bright stellar emission and from FSRQs, respectively, are not expected to be polarized, we have studied the depolarizing effect that these two contributions can have on the polarization degree of the non-thermal synchrotron emission of the relativistic jet (Section \ref{sec:3.2}). We have also studied the variability of the polarization degree and searched for rotations in the evolution of the polarization angle (Section~\ref{sec:3.3}).

\subsection{Variability of the polarization}\label{sec:3.1}
With the aim of studying and quantifying the variability of the polarization, we have tested the statistical distribution that provides a best fit for the polarization degree. Following the approach from \cite{blinov2016a}, we have evaluated the probability density function (PDF) with a Beta distribution defined as
\begin{equation}
\text{PDF$_{\text{Beta}}$}(x,\alpha,\beta)=\frac{x^{\alpha-1}(1-x)^{\beta-1}}{B(\alpha,\beta)},
\label{eq:beta_dist}
\end{equation}
where $x$ represents the polarization degree, $\alpha$ and $\beta$ (with $\alpha,\beta>0$) are the parameters that define the distribution and $B(\alpha,\beta)$ is the Beta function. {This consideration has been used by other authors in the past \cite[for instance,][]{hovatta2016}, with a good agreement between the data and the Beta distribution. In order to assess the reliability and success of this function when describing our data, we have applied a $\chi^{2}$ test \citep{pearson1900} to account for the goodness of the fit, considering values of $\chi^{2}/\text{d.o.f.} \lesssim 1$ as good fits. As a comparison, we have also tested the performance of the commonly used Gaussian distribution, typically assumed for the estimation of several parameters in blazar studies, e.g. the fractional variability, $F_{var}$.}


Assuming that the Beta distribution is able to reproduce the measured polarization degree, the mean value can be expressed as a function of $\alpha$ and $\beta$,
\begin{equation}
P_{mean}=\frac{\alpha}{\alpha+\beta}.
\label{eq:beta_dist_mean}
\end{equation}
In addition, the parameters of the distribution can also be used for quantifying the variability displayed by the polarization degree, providing an estimation of the so-called modulation index
\begin{equation}
m_{int}=\frac{\sqrt{\sigma^{2}}}{P_{mean}},
\label{eq:beta_dist_modulation_index}
\end{equation}
where $\sigma^{2}$ is the variance of the Beta distribution \citep{hovatta2016}. This quantity can be obtained from $\alpha$ and $\beta$ as
\begin{equation}
\sigma^{2}=\frac{\alpha \beta}{(\alpha+\beta)^{2}(\alpha+\beta+1)}.
\label{eq:beta_dist_variance}
\end{equation}
The modulation index $m_{int}$ is analogous to the fractional variability, $F_{var}$, commonly used in blazar studies, that quantifies the intensity of the variability. As a crosscheck, we have also estimated the $F_{var}$ following the prescription from \cite{vaughan2003}, as
\begin{equation}
F_{var}=\sqrt{\frac{S^2-\langle\sigma^2_{err}\rangle}{\langle x \rangle^2}}.
\label{eq:fractional_variability_equation}
\end{equation}
$S^{2}$ represents the variance of the data sample, $\langle \sigma^2_{err} \rangle$ is the mean square error and $\langle x \rangle$ corresponds to the mean value of the sample. The uncertainty of this quantity can be estimated following Equation~(B2) from \cite{vaughan2003}. {Notice that the $F_{var}$ is associated with a Gaussian distribution. Therefore, it is used only for comparative purposes. The results of the behaviour of the polarized emission using the tools presented here will be discussed in Section~\ref{sec:results}.} 

\subsection{Depolarizing effect of non-polarized components}\label{sec:3.2}
Studies such as \cite{sosa2017} or \cite{blinov2021} have claimed in the past that the host galaxy can have a depolarizing effect in the measured polarization degree of the synchrotron emission. This effect can be especially important in those sources with a bright host galaxy (e.g. those presented as galaxy-dominated blazars in OS22), with an optical emission comparable to that from the relativistic jet. To account for this effect in the three galaxy-dominated sources in our sample (H~1426+428, Mkn~501 and 1ES~2344+514) we have made use of the estimation of the host galaxy performed in OS22. This estimation is based on the decomposition of all the spectral dataset of the blazar using the minimum number of meaningful components, associated with the different contributions to the emission, that are able to successfully reproduce and model the variability displayed by the source by using the Non-Negative Matrix Factorization \citep[NMF,][]{paatero1994}. With this approach we have been able to estimate the overall contributions of both the jet and the host galaxy and separate their emission. With this estimate of the host galaxy, we can subtract it in order to correct the measured polarization degree from the stellar contribution as
\begin{equation}
P_{intr}(t) \ [\%] = \frac{P_{obs}(t) \ [\%]}{1-\frac{I_{\text{host galaxy}}}{I_{\text{total}}(t)}}.
\label{eq:depolarization_host_galaxy}
\end{equation}

This correction can also be applied to FSRQs, where the BLR is also expected to be unpolarized. Making use of the BLR component extracted from the NMF reconstruction from OS22, and again following Equation (\ref{eq:depolarization_host_galaxy}), we can obtain an estimation of the intrinsic polarization degree of the synchrotron emission, higher than the observed polarization fraction. {Notice that by BLR contribution we mean not only the broad emission line component but also the thermal continuum likely coming from the accretion disc as observed in radio-quiet QSOs \citep{vandenberk2001}. Hereafter we will refer to this contribution as the {BLR+AD} component.}

\subsection{EVPA rotations}\label{sec:3.3}
The variability of the polarization is not only restricted to changes in the polarization degree. Changes of the polarization angle orientation are also observed in the temporal evolution of blazars. These changes are often seen with an erratic behaviour. However, in some occasions, smooth and continuous swings in the variability of the EVPA are detected \citep[see for instance][]{blinov2015,blinov2016a,blinov2016b,kielhmann2017}. These events, typically identified as EVPA rotations or swings, have been one of the ways to study the behaviour of the polarization and magnetic field of the relativistic jets in these objects. 

\subsubsection{Smooth rotations}\label{sec:smooth_rotations}
There is no standard definition of an EVPA rotation in the framework of blazars. Therefore, we have followed a similar definition to that used by the RoboPol programme, detailed in \cite{blinov2015}. The main characteristics that an EVPA variation must fulfill to be considered a rotation are the following:

\begin{itemize}
    \item A minimum variation of the polarization angle $\Delta \theta \geq 90^{\circ}$.
    
    \item In order to exclude possible false identifications, the EVPA swing must contain at least 4 observations.
    
    \item Owing to the fast variability displayed by blazars, consecutive points of a rotation must be separated in time by $\Delta t \leq 30$~days.
    
    \item The start and end of the rotation must be accompanied by a change of slope $\Delta \theta_{i} / \Delta t_{i}$ w.r.t. previous observations. Following the prescription from \cite{blinov2015}, we used a factor 5 change, or a change of sign, to define the start and end of the rotation.
    
\end{itemize}

Furthermore, due to the fast variability and fluctuations observed in the emission of blazars, we do not reject rotations in which a point slightly fluctuates from the general trend of the EVPA swing, allowing one measurement to deviate <5$^{\circ}$ if the following measurement continues the general trend of the rotation. An example of these rotations is represented in the top panel of Figure \ref{fig:rotation_examples}. 

\subsubsection{Slow and non-smooth rotations}\label{sec:nonsmooth_rotations}
In addition to the aforementioned smooth rotations, here we have also considered those rotations showing a non-smooth, and typically much slower change of the EVPA. A visual inspection of these non-smooth swings reveals that they tend to have a higher angle variation and duration than the smooth swings, leading to the appearance of several fluctuations and fast variability in shorter time scales during the development of the angle rotation. 

\begin{figure}
\centering
\subfigure{\includegraphics[width=0.9\columnwidth]{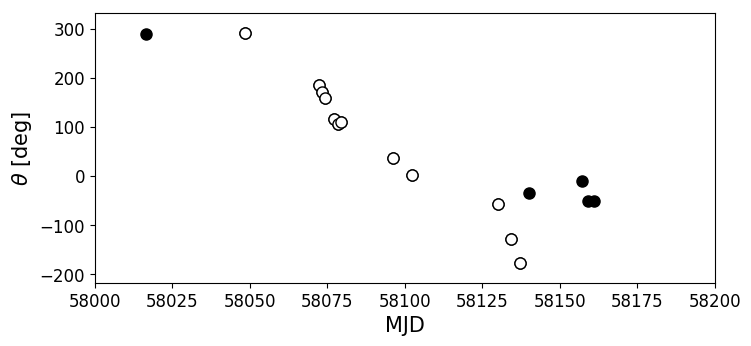}}
\vspace{-0.3cm}
\subfigure{\includegraphics[width=0.9\columnwidth]{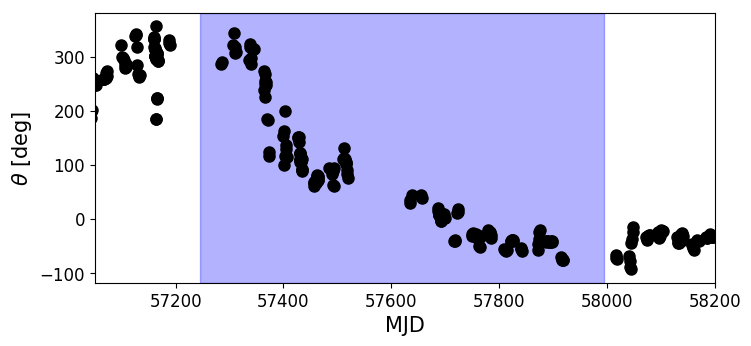}}
\caption{Examples of the two types of rotations considered here. \textit{Top:} Smooth rotation detected for AO~0235+164. Empty markers represent the observations during the rotation. \textit{Bottom:} Slow non-smooth rotation observed for OJ~287. The EVPA swing occurs during the period contained in the blue contour.}
\label{fig:rotation_examples}
\end{figure}

In order to identify these non-smooth swings, we use the same definition introduced above, without the condition demanding a continuous change of the polarization angle during the rotation (see bottom panel of Figure \ref{fig:rotation_examples}). Prior to the identification, we also apply a cubic spline interpolation on the binned EVPA light curve, with bins of 60-100~days depending on the source, to remove the variability in short time scales of the polarization angle. This smoothing leads to an easier identification of these non-smooth EVPA rotations. 

\section{Results}\label{sec:results}

\subsection{General behaviour of the polarization}\label{sec:general_behaviour}
\begin{figure*}
\centering
    \includegraphics[width=0.5\textwidth]{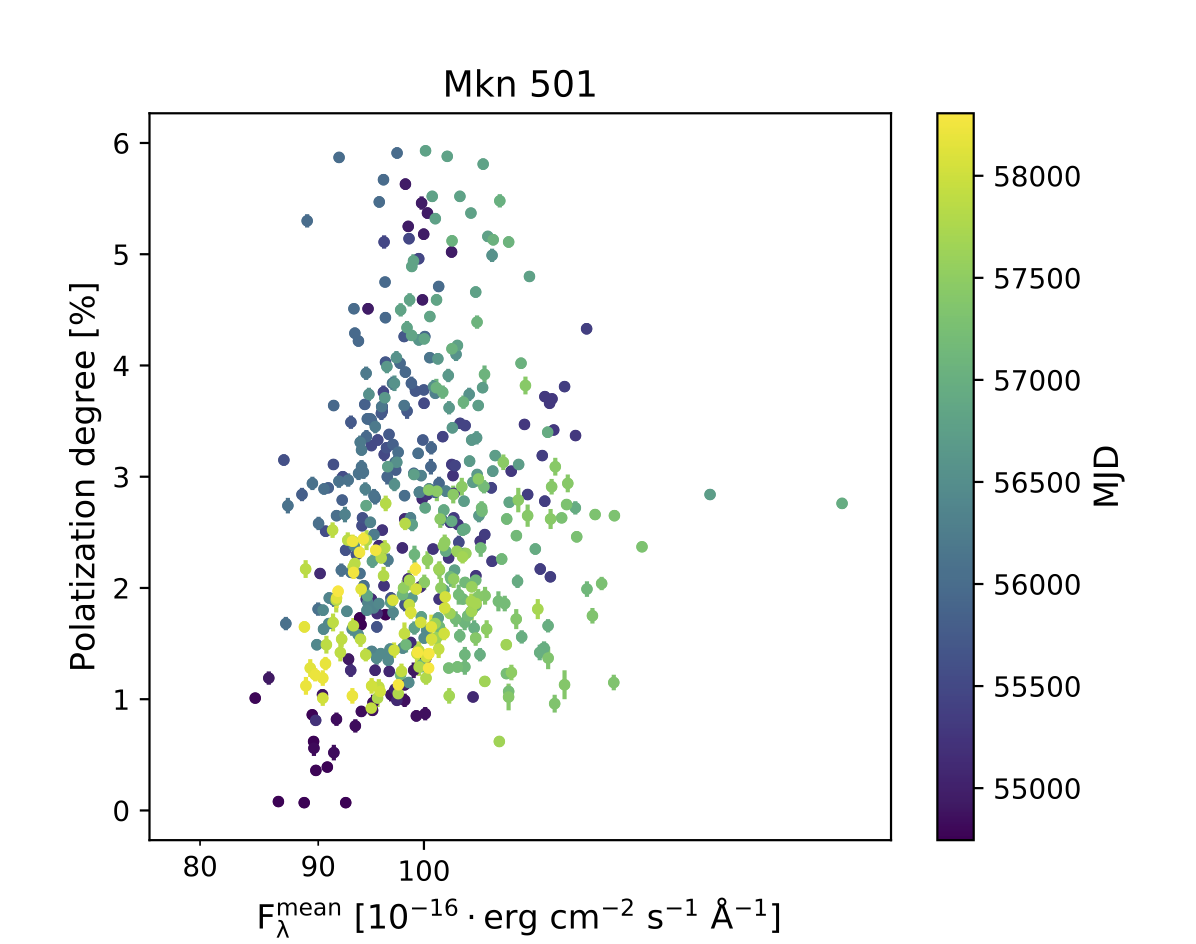}\hfill
    \includegraphics[width=0.5\textwidth]{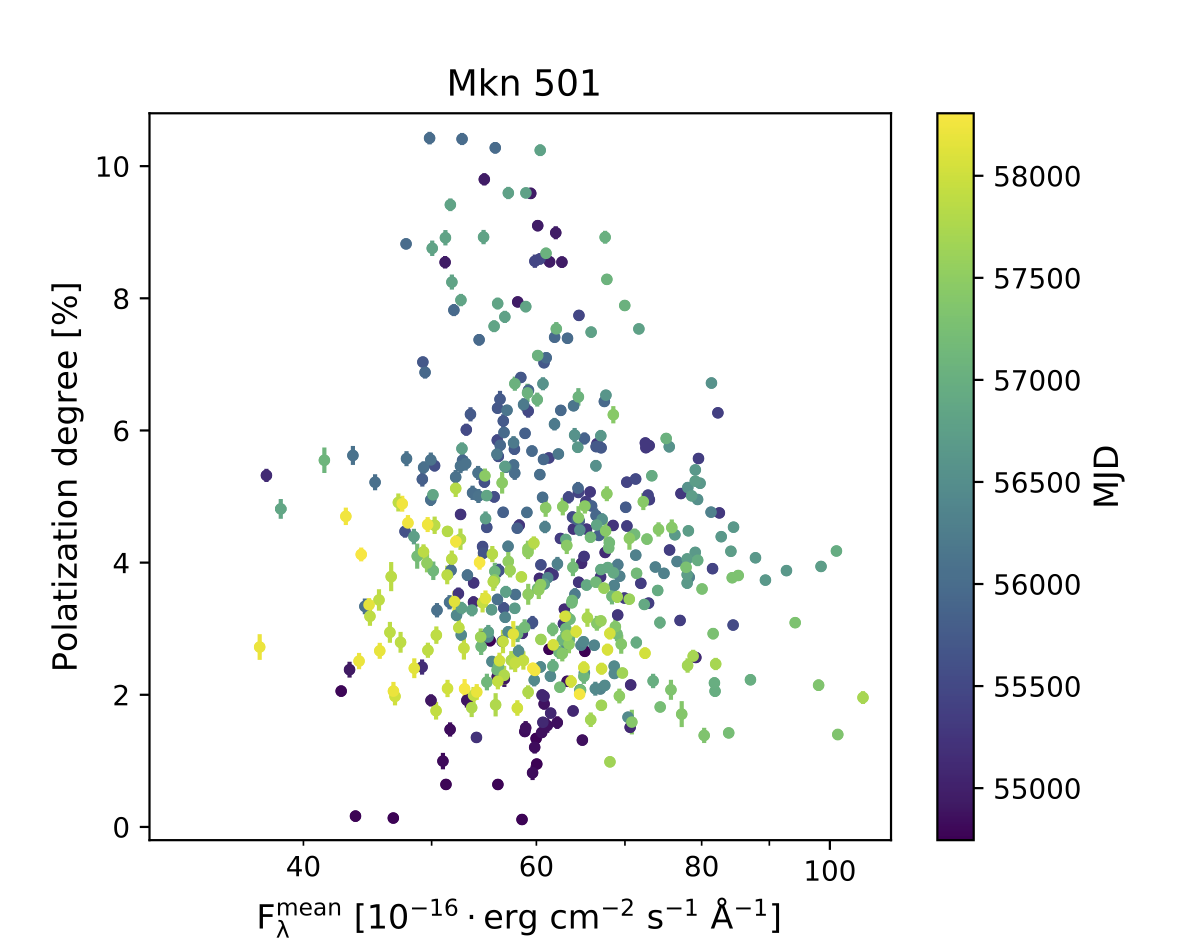}\hfill
    \caption{Polarization-flux diagrams for the galaxy-dominated blazar Mkn~501. The flux value is estimated from the mean flux of each total optical spectrum. \textit{Left}: Before correcting by the host galaxy. \textit{Right}: After correcting by the host galaxy.}
    \label{fig:polarization_galdom_mkn501}
\end{figure*}

\begin{figure}
	\centering
	\includegraphics[width=\columnwidth]{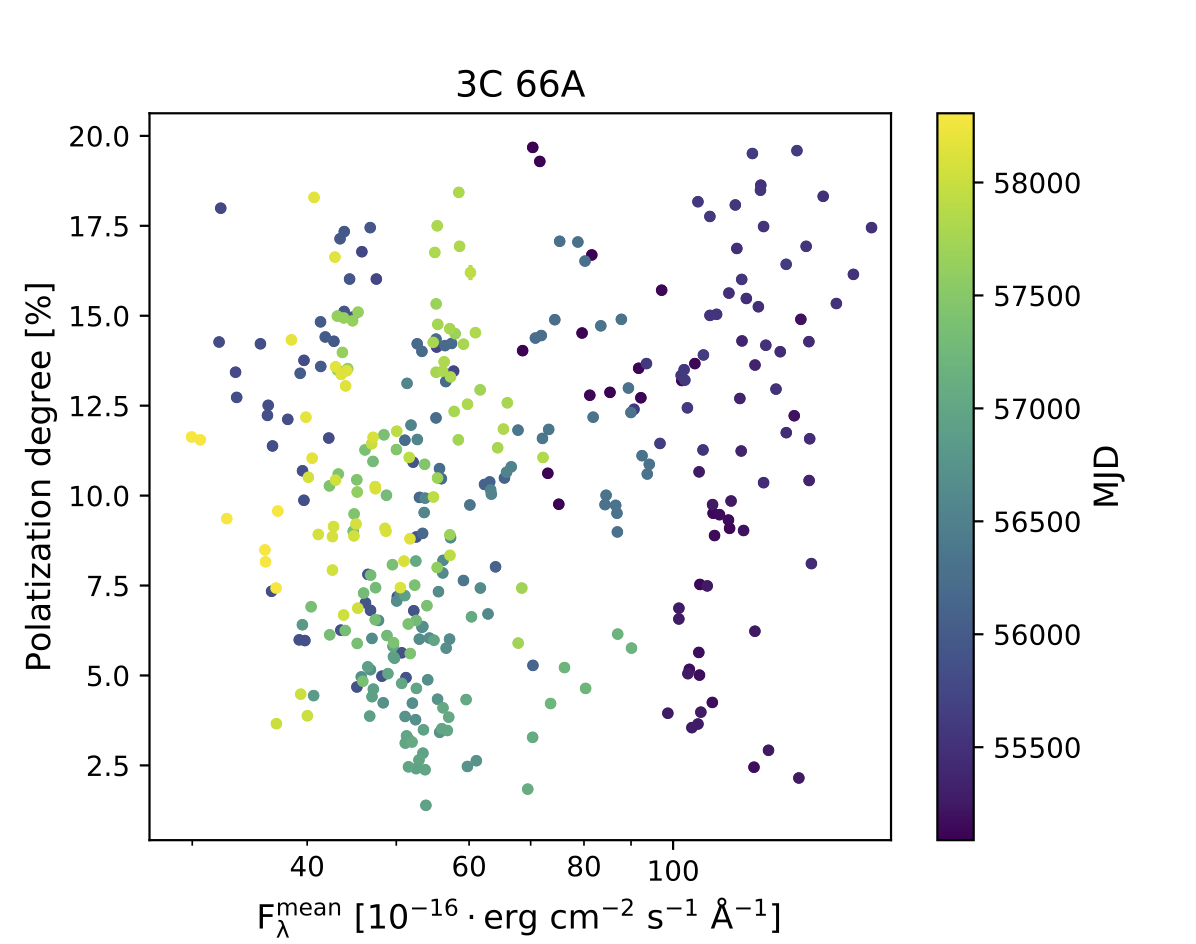}\hfill
	\caption{Polarization-flux diagram for the BL Lac object 3C~66A. The flux value is estimated from the mean flux of each total optical spectrum.}
	\label{fig:polarization_bllac_3c66a}
\end{figure}

\begin{figure*}
\centering
    \includegraphics[width=0.5\textwidth]{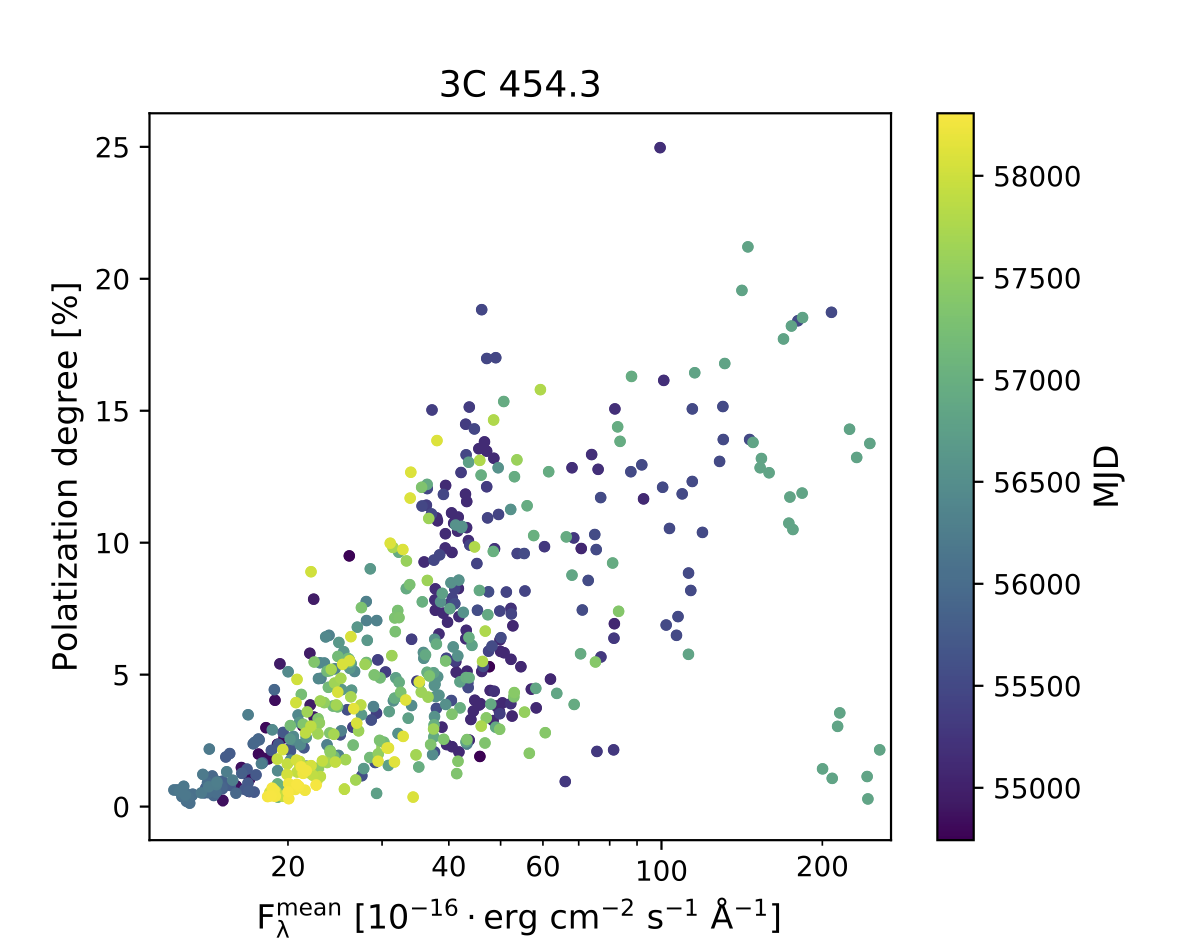}\hfill
    \includegraphics[width=0.5\textwidth]{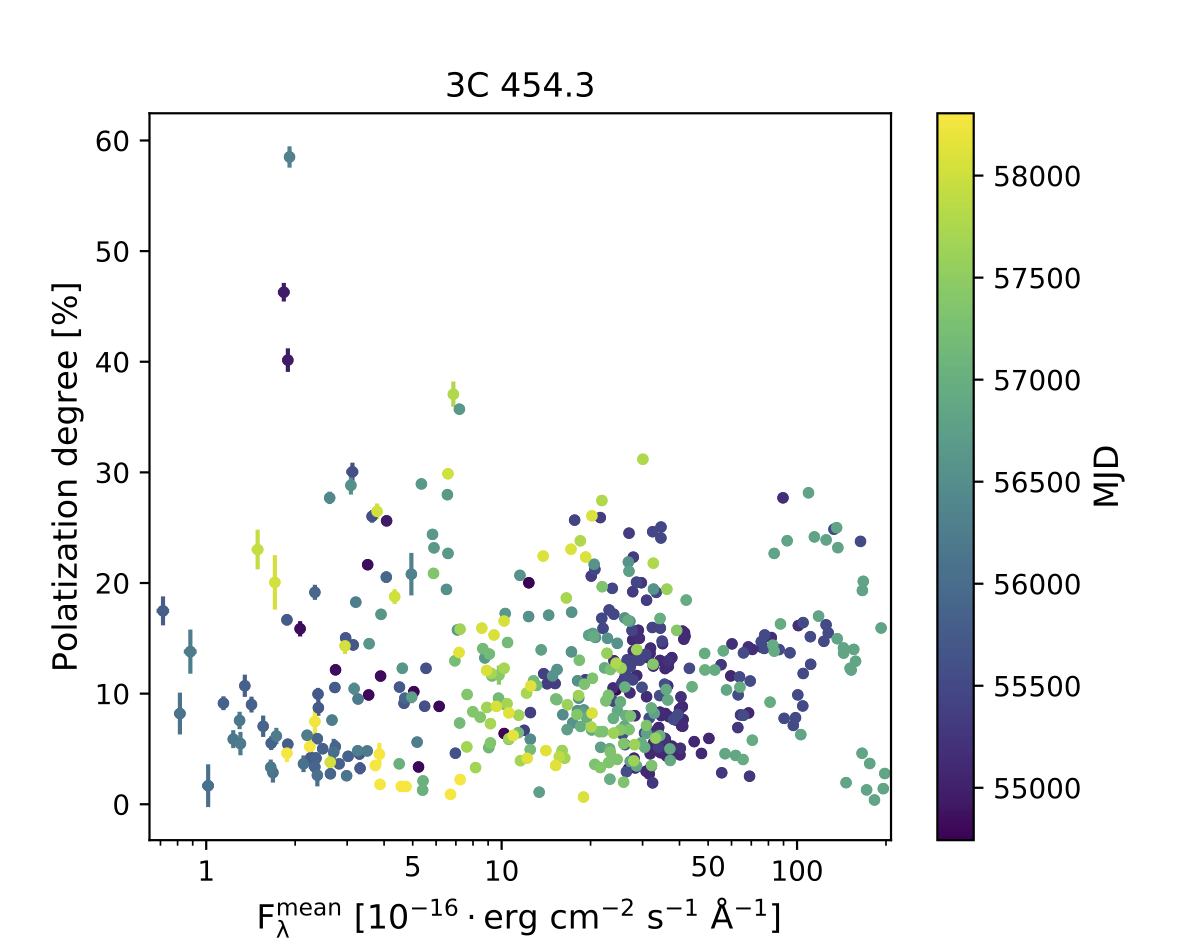}\hfill
    \caption{Polarization-flux diagrams for the FSRQ blazar 3C~454.3. The flux value is estimated from the mean flux of each total optical spectrum. \textit{Left}: Before correcting by the {BLR+AD}. \textit{Right}: After correcting by the {BLR+AD}.}
    \label{fig:polarization_fsrq_3c454}
\end{figure*}

{The long-term evolution of the polarization and its relation with the total optical flux {was} studied from the polarization-flux diagrams. As an illustration we represent an example for each class: the galaxy-dominated blazar Mkn~501, the BL Lac object 3C~66A and the FSRQ 3C~454.3, in Figures \ref{fig:polarization_galdom_mkn501}, \ref{fig:polarization_bllac_3c66a} and \ref{fig:polarization_fsrq_3c454}, respectively. The figures corresponding to the rest of the sources are included as online material (Figures~\ref{fig:polarization_galdom_online} to \ref{fig:polarization_fsrqs_online}).} 
{The polarization-flux diagrams prior to the depolarizing correction for galaxy-dominated blazars and FSRQs are also displayed the in left panels of Figures \ref{fig:polarization_galdom_mkn501} and \ref{fig:polarization_fsrq_3c454}, as well as the corresponding online figures for each blazar.} The {galaxy-dominated blazars show} a very low polarization in comparison to BL Lacs and FSRQs, with values {of the polarization degree before host galaxy correction} <6\%, and average values of $\sim$3\%. {A linear correlation coefficient of $r \sim 0.55$, indicates} a mild correlation between the polarization degree and the total optical emission.

\begin{figure}
	\centering
	\includegraphics[width=\columnwidth]{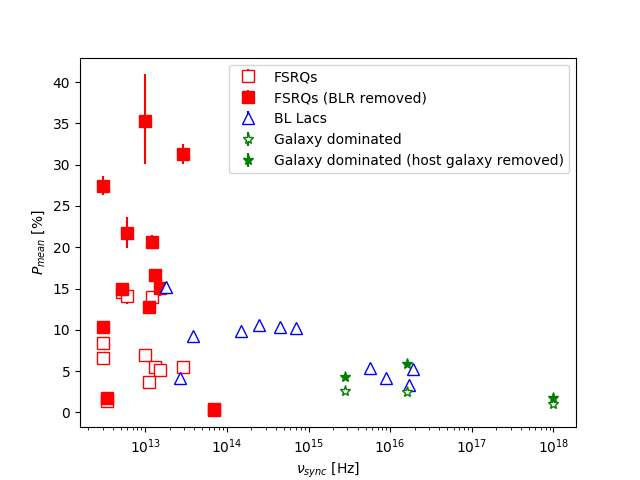}
	\caption{Mean polarization degree vs. the frequency of the synchrotron peak $\nu_{sync}$. Open and filled green stars represent the galaxy-dominated sources with and without the contribution of the host galaxy, respectively. Blue triangles correspond to the BL Lac objects. Open and filled red squares correspond to the FSRQs, with and without the {BLR+AD} contribution, respectively (see Section \ref{sec:depolarization}).}
	\label{fig:polariation_degree_frequency}
\end{figure}

Concerning the BL Lac objects, we do not observe any relation between their polarization and their total optical emission. This can be seen for example in Figure \ref{fig:polarization_bllac_3c66a}. They tend to have a higher polarization fraction than those BL Lacs dominated by the emission of the stellar population, with typical mean percentages of 5-10\%. Additionally, through a visual inspection of Figure \ref{fig:polariation_degree_frequency} where the mean polarization degree is represented w.r.t. the frequency of the synchrotron peak of the SED, we observe that for BL Lacs with a lower peaking frequency, a higher polarization degree is measured, in comparison to those with a high $\nu_{sync}$, with very low values of the polarization fraction. {This is also reflected in the correlation coefficient between these two quantities estimated for the BL Lac population, with a value of $r=-0.65$, $\text{p-value}=0.03$.}

On the contrary, FSRQs show different types of behaviour before the {BLR+AD} correction, as it can be observed for example in the left panel of Figure~\ref{fig:polarization_fsrq_3c454} for {3C~454.3} before correcting the effect of the {BLR+AD}. Five of the sources of this type show hints or {evidence} of a correlated evolution of the polarized and total emission, with correlation coefficients $r \geq 0.50$: PKS~0420-014, OJ~248, B2~1633+38, 3C~454.3 {and CTA~102}. This behaviour was reported already in the past for B2~1633+38 by \cite{raiteri2012}. Remarkably, for one of the FSRQs of the sample, 3C~345, the polarization degree displays an anticorrelation with the total optical flux. The correlation coefficient for this blazar {is} $r=-0.49$, $\text{p-value}=10^{-3}$. The remaining six FSRQs do not show any significant correlation or behaviour in their long-term polarization variability. Regarding the relation between the mean polarization and the location of the synchrotron peak of the SED, FSRQs {and cover a higher range of mean polarization fractions than BL Lac objects with a high $\nu_{sync}$ and galaxy-dominated blazars ($\nu_{sync}$ > 10$^{15}$~Hz), as shown in Figure~\ref{fig:polariation_degree_frequency}.} 

\begin{figure}
	\centering
	\includegraphics[width=\columnwidth]{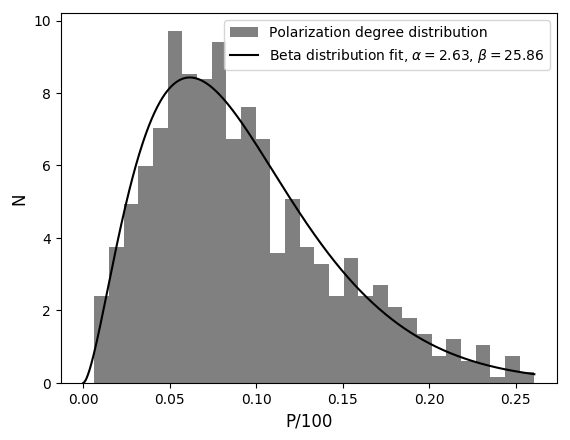}
	\caption{Beta distribution of BL Lacertae for its normalized polarization degree. The grey shaded area represents the distribution of the polarization degree measurements and the black line corresponds to the best fit of the Beta distribution. The fitted values of $\alpha$ and $\beta$ are also shown in the legend of the figure.}
	\label{beta_distribution_example}
\end{figure}

In addition to the long-term variability of the polarization degree, we have evaluated the characteristics of the polarized emission of this blazar sample. We have tested the distribution that describes best the measured polarization degree through a {$\chi^{2}$} test. This test was used to compare the goodness of the fit of the Beta distribution introduced in Section \ref{sec:methodology} and a Gaussian distribution, using the best fit to the data of each of them as the null hypothesis. {We obtain typical values of $\chi^{2}/\text{d.o.f.} \lesssim 1$ and p-values<0.05 for 21 of the 26 blazars studied here. Therefore, the Beta distribution leads to a good agreement with the data for most of the sources of the sample. An example of a Beta PDF fit for BL~Lacertae is presented in Figure \ref{beta_distribution_example}. For the remaining five sources, the Beta distribution is not able to accurately describe the distribution of their polarization degree. It is important to highlight that for three of these five blazars, the {low} number of observations ($N_{obs} \leqslant 47$) is most likely affecting the result of the test. The last two targets for which the Beta distribution results in a poor fit to the data are 3C~273 and OJ~248. These two blazars are characterised by an almost unpolarized emission, with the exception of a bright flare followed by an increase of the polarization degree for the latter. The fact that the majority of the population is well represented by a Beta distribution is in line with the results presented by \cite{blinov2016a}. On the other hand, the Gaussian distribution fits lead to values of $\chi^{2}/\text{d.o.f.} > 1$, proving that it is not suitable for describing the distribution of the polarization degree of blazars.} 

\subsection{Depolarizing effect of the host galaxy and the {BLR+AD}}\label{sec:depolarization}
We have considered the effect of the host galaxy in the measured polarization degree for those blazars showing a bright stellar emission, namely H~1426+428, Mkn~501 and 1ES~2344+514. Following Equation (\ref{eq:depolarization_host_galaxy}) and considering the estimation of the host galaxy performed in OS22, we are able to calculate the intrinsic polarization degree of the emission of the jet. We observe that after subtracting the contribution of the host galaxy, the polarization degree of these sources shows an increase of approximately a factor 2 w.r.t. the observed measurements. Moreover, the corrected polarization degree does not show the mildly correlated behaviour with the total flux commented in Section \ref{sec:general_behaviour}. On the other hand, it now presents the same behaviour as the rest of the BL Lac sample, with no apparent trend between the total optical flux and the polarization. This behaviour can be observed in {the right panel of} Figure~\ref{fig:polarization_galdom_mkn501} {for the case of Mkn~501, and in Figure \ref{fig:polarization_galdom_online} for H~1426+428 and 1ES~2344+514}.


This correction has also been applied to FSRQs, where the contribution of the {BLR+AD} can also have a depolarizing effect. Depending on the different relative contribution of the {BLR+AD} for each FSRQ, the correction ranges from almost an unchanged polarization degree for those sources with a faint {BLR+AD}, up to a factor $\gtrsim$3 for the brightest {BLR+AD} (e.g. PKS~1510-089). The corrected polarization degree for these sources w.r.t. the flux is presented in {the right panel of} Figure~\ref{fig:polarization_fsrq_3c454} {for the case of 3C~454.3, and in Figure \ref{fig:polarization_fsrqs_online} for the rest of the FSRQs}. It is also remarkable that after this correction, we observe polarization degrees as high as 50-70\% for example for Ton~599, PKS~1222+216 or 3C~454.3.

Regarding the behaviour of the polarization degree with the total flux, as for the galaxy-dominated blazars, the trends observed between these two quantities are no longer observable after the correction, with the exception of OJ~248 and 3C~345. The former presents a positive correlation with a linear correlation coefficient $r$=0.76 (p-value=5.4$\times$10$^{-44}$). \cite{carnerero2015} and \cite{raiteri2021} also reported the observed correlation for this source, attributing the variability to a turbulent magnetic field component. The latter displays an anticorrelated behaviour, with a coefficient $r$=-0.80 (p-value=1.4$\times$10$^{-6}$).

\subsection{Polarization variability}\label{sec:polarization_variability}
\begin{figure*}
\subfigure{\includegraphics[width=0.49\textwidth]{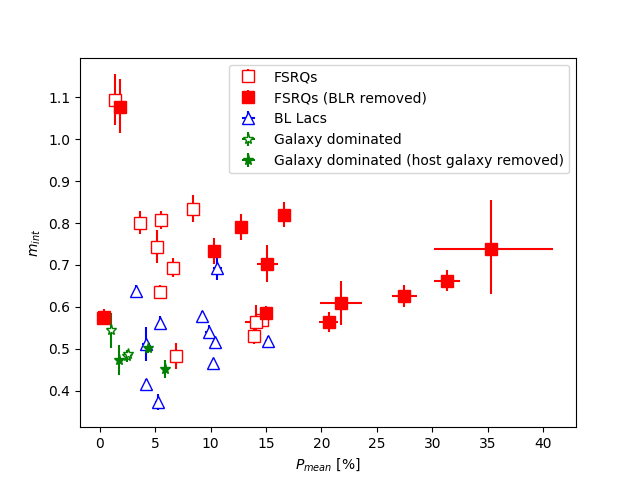}}
\subfigure{\includegraphics[width=0.49\textwidth]{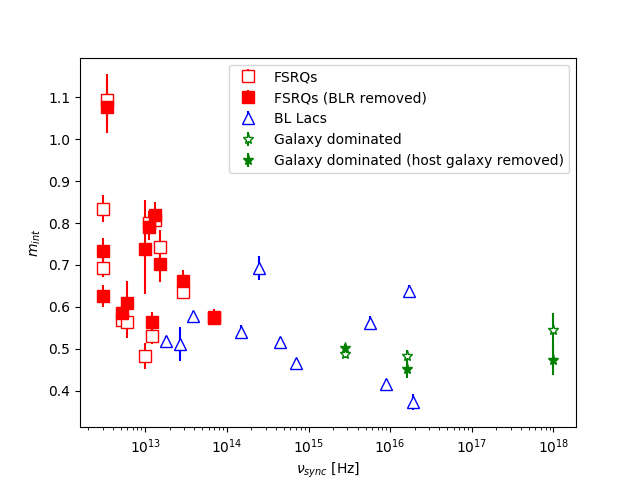}}
\caption{Intrinsic modulation index of the polarization degree. \textit{Left:} $m_{int}$ vs. the mean polarization degree. \textit{Right:} $m_{int}$ vs $\nu_{sync}$. Open and filled green stars represent the galaxy-dominated sources with and without the contribution of the host galaxy, respectively. Blue triangles correspond to the BL Lac objects. Open and filled red squares correspond to the FSRQs, with and without the {BLR+AD} contribution, respectively.}
\label{fig:intrinsic_modulation_index_polarization}
\end{figure*}

We have evaluated the variability of the polarization. As a first step, we have compared the intensity of the variability derived from the modulation index $m_{int}$ and the fractional variability $F_{var}$ defined by Equations (\ref{eq:beta_dist_modulation_index}) and (\ref{eq:fractional_variability_equation}), respectively. We find that both measurements are compatible within errors. Therefore, owing to the good agreement of the Beta distribution with the data, and previous works using $m_{int}$ to define the variability of the polarization \citep[e.g.][]{hovatta2016}, hereafter we will refer to the results of the intrinsic modulation index. The modulation index for each source is represented in {the left} panel of Figure \ref{fig:intrinsic_modulation_index_polarization} w.r.t. {$P_{mean}$}. In the right panel we also present $m_{int}$ with the frequency of the synchrotron peak of the SED.

This analysis clearly shows a higher variability coming from FSRQs in comparison to BL Lac and galaxy-dominated sources. The mean modulation index for BL Lacs and galaxy-dominated sources was estimated to be $m_{int}=0.53 \pm 0.08$ and $m_{int}=0.50 \pm 0.03$, respectively. On the other hand, FSRQs yield a higher value $m_{int}=0.70 \pm 0.13$. This separation is also visible in the right panel of Figure~\ref{fig:intrinsic_modulation_index_polarization}, where sources with a lower synchrotron peak display a higher $m_{int}$ value. This trend leads to a correlation coefficient $r$=-0.52 (p-value=0.007). The left panel of Figure \ref{fig:intrinsic_modulation_index_polarization} does not reveal any evident difference in the mean polarization degree {prior to the host galaxy and BLR+AD correction} between BL Lacs and FSRQs with $P_{mean}=(7.97 \pm 0.04)$\% and $P_{mean}=(7.16 \pm 0.05)$\%, respectively, with only the galaxy-dominated blazars presenting a significantly lower mean polarization degree $P_{mean}=(2.02 \pm 0.01)$\%. However, after accounting for the depolarization produced by the {BLR+AD}, FSRQs appear to be more polarized than BL~Lac objects, with a mean intrinsic polarization degree of $P_{mean}=(17.37 \pm 0.10)$\%. In addition, the sources presenting a bright host galaxy show an increase of their mean polarization degree of a factor $\sim$2 after correcting by the contribution of the host galaxy, $P_{mean}=(3.95 \pm 0.02)$\%. In fact, after accounting for the depolarizing effect, we observe an anticorrelation between the mean polarization and the value of $\nu_{sync}$, with $r$=-0.57 (p-value=0.002), as shown in Figure \ref{fig:polariation_degree_frequency}. Despite this increase of the mean polarization for these blazar types, no significant change in the intrinsic modulation index is observed after the correction. Compatible results in this regard have also been reported by \cite{angelakis2016}. A similar {synchrotron peak frequency dependence} of the polarization degree and its variability has been reported by \cite{smith1996}. We note that for some sources, the frequency of the synchrotron peak was observed to change depending on the emission state \citep[e.g. Mkn~501, see][]{acciari2020}. However, as observed from Figure \ref{fig:polariation_degree_frequency}, this trend is clearly visible despite small shifts of $\nu_{sync}$. 

\begin{figure}
	\includegraphics[width=\columnwidth]{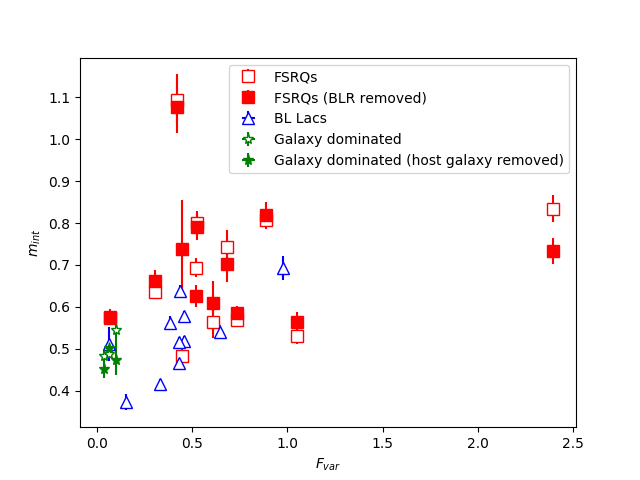}
	\caption{Intrinsic modulation of the polarization vs. the optical fractional variability. Open and filled green stars represent the galaxy-dominated sources with and without the contribution of the host galaxy, respectively. Blue triangles correspond to the BL Lac objects. Open and filled red squares correspond to the FSRQs, with and without the {BLR+AD} contribution, respectively.}
	\label{fig:intrinsic_modulation_vs_fvar}
\end{figure}

In addition to this, we have also evaluated the relation between the intrinsic modulation index of the polarization degree and the fractional variability of the total optical flux, extracted from Table 2 of OS22, and represented in Figure \ref{fig:intrinsic_modulation_vs_fvar}. This figure clearly shows that sources with a higher variability in the optical band also present a higher variability and $m_{int}$ for the polarization degree, as expected. 

\begin{table*}
\begin{center}
\caption{Sample of blazars from the Steward Observatory monitoring programme included in the polarization analysis. Columns: (1) Target name. (2) Source type. (3) Number of polarization measurements. (4) Mean observing cadence. (5) Mean observed polarization degree. (6) Intrinsic modulation index of the polarization. (7) Preferential EVPA direction. (8) Linear correlation coefficient between the total flux and the polarization degree. (9) Number of displayed smooth and fast rotations. (10) Total spectral index range. (11) Polarized spectral index range. \label{tab:polarization_results}}
\begin{adjustbox}{max width=\textwidth}
\hspace{-0.4cm}
\begin{tabular}{ccccccccccc} \hline \hline
(1) & (2) & (3) & (4) & (5) & (6) & (7) & (8) & (9) & (10) & (11) \\ 
Source & Type  & $N_{obs}$ & $\Delta t_{mean}$ [days] & $P_{mean}$ [\%] & $m_{int}$ & $\theta_{pref}$ [$^{\circ}$]$^{a}$ & r$^{b}$ & $N_{rot}$ & $\alpha_{tot}$ & $\alpha_{pol}$ \\ \hline \hline
H 1426+428 & Galaxy dominated  & 47 & 12.8 & $1.03 \pm 0.08$ & $0.54\pm0.04$ & --$^{c}$  & 0.55 & 0 & 1.3--1.4 & -0.3--2.5 \\ \hline
Mkn 501 & Galaxy dominated  & 624 & 5.0 & $2.56 \pm 0.05$ & $0.49\pm0.01$ & -56.3 $\pm$ 6.3 & 0.14 & 0 & 1.3--1.5 & 0.5--2.5 \\ \hline
1ES 2344+514 & Galaxy dominated  & 246 & 14.5 & $2.46 \pm 0.07$ & $0.48\pm0.02$ & -46.9 $\pm$ 6.1 & 0.54 & 0 & 0.8--1.0 & 0.0--2.5 \\ \hline \hline
3C 66A & BL Lac &  477 & 6.1 & $10.20 \pm 0.22$ & $0.70\pm0.01$ & 25 $\pm$ 13, -20 $\pm$ 12$^{d}$ & 0.23 &  0 & 0.7--1.7 & 0.5--2.0 \\ \hline
AO 0235+164 & BL Lac  & 277 & 13.2 & $10.58 \pm 0.43$ & $0.69\pm0.03$ & Uniform & 0.31  &  9 & -1.1--1.5 & -1.3--2.2 \\ \hline
S5 0716+714 & BL Lac  & 325 & 7.1 & $9.80 \pm 0.29$ & $0.54\pm0.02$ & Uniform & -0.01 &  8 & 0.4--1.3 & 0.5--1.9 \\ \hline
PKS 0735+178 & BL Lac  & 39 & 4.6 & $4.17 \pm 0.33$ & $0.51\pm0.04$ & --$^{c}$ & -0.12 &  0 & 0.6--0.9 & 0.5--1.8\\ \hline
OJ 287 & BL Lac  & 708 & 3.7 & $15.18 \pm 0.30$ & $0.52\pm0.01$ & -41.0 $\pm$ 7.2 & 0.19 & 1 & 0.2--1.7 & 0.2--1.9\\ \hline
Mkn 421 & BL Lac  & 859 & 3.7 & $3.26 \pm 0.07$ & $0.64\pm0.01$ & -15.4 $\pm$ 6.9 & 0.30 & 12 & 0.7--2.2 & 0.7--2.8 \\ \hline
W Comae & BL Lac  & 454 & 6.0 & $10.39 \pm 0.25$ & $0.52\pm0.01$ & 71.2 $\pm$ 6.6 & -0.10 & 5 & 0.1--1.3 & 0.0--1.6 \\ \hline
H 1219+305 & BL Lac  & 47 & 6.6 & $5.22 \pm 0.28$ & $0.37\pm0.02$ & --$^{c}$ & -0.20 & 0 & 0.6--1.4 & 0.5--2.5 \\ \hline
1ES 1959+650 & BL Lac  & 202 & 17.0 & $4.17 \pm 0.12$ & $0.42\pm0.02$ & -35.9 $\pm$ 5.9 & -0.06  & 0 & 0.4--1.6 & 0.6--2.1 \\ \hline
PKS 2155-304 & BL Lac  & 425 & 7.1 & $5.42 \pm 0.14$ & $0.56\pm0.01$ & 52.4 $\pm$ 6.8  & 0.22 & 3 & 1.0--2.1 & 1.0--2.5 \\ \hline
BL Lacertae & BL Lac  & 788 & 4.1 & $9.23 \pm 0.19$ & $0.58\pm0.01$ & 17.4 $\pm$ 6.1 & -0.23 & 14 & -0.3--1.3 & -0.3--2.0 \\ \hline \hline
PKS 0420-014 & FSRQ  & 64 & 13.2 & $14.10 \pm 1.00$ & $0.56\pm0.04$ & Uniform & 0.56 & 1 & 0.5--1.7 & 0.2--1.5 \\ \hline
PKS 0736+017 & FSRQ  & 172 & 5.0 & $5.16 \pm 0.28$ & $0.74\pm0.04$ & Uniform & 0.34 & 9 & 0.5--2.3 & 0.0--2.5 \\ \hline
OJ 248 & FSRQ  & 266 & 10.9 & $1.40 \pm 0.08$ & $1.09\pm0.06$ & Uniform & 0.76 & 3 & 1.3--2.3 & 1.0--2.5 \\ \hline
Ton 599 & FSRQ  & 260 & 5.7 & $13.94 \pm 0.46$ & $0.53\pm0.02$ & Uniform & -0.05 & 7 & 0.9--2.2 & 0.8--2.5 \\ \hline
PKS 1222+216 & FSRQ & 558 & 4.1 & $5.46 \pm 0.14$ & $0.64\pm0.02$ & 1.6 $\pm$ 6.1 & 0.37 & 4 & 2.2--3.3 & 0.3--2.1\\ \hline
3C 273 & FSRQ & 396 & 6.7 & $0.24 \pm 0.01$ & $0.57\pm0.02$ & 63.0 $\pm$ 6.3 & 0.03 &  0 & 2.4--3.5 & 1.5--4.0\\ \hline
3C 279 & FSRQ & 660 & 3.9 & $14.59 \pm 0.32$ & $0.57\pm0.01$ & 46.7 $\pm$ 6.7 & 0.16 & 12 & 0.5--1.9 & 0.2--2.0 \\ \hline
PKS 1510-089 & FSRQ  & 471 & 6.4 & $3.63 \pm 0.12$ & $0.80\pm0.03$ & Uniform & 0.35 &  9 & 1.1--3.0 & 0.3--3.1 \\ \hline
B2 1633+38 & FSRQ  & 441 & 7.6 & $6.64 \pm 0.21$ & $0.69\pm0.02$ & Uniform & 0.70 & 5 & 1.3--3.0 & 0.1--2.2 \\ \hline
3C 345 & FSRQ  & 52 & 28.1 & $6.90 \pm 0.46$ & $0.48\pm0.03$ & 64.3 $\pm$ 6.6 & -0.49 & 1 & 1.5--2.7 & 0.5--2.7 \\ \hline
CTA 102 & FSRQ  & 410 & 12.4 & $8.39 \pm 0.33$ & $0.83\pm0.03$ & Uniform & 0.28 &  11 & 0.8--3.0 & 0.4--3.0 \\ \hline
3C 454.3 & FSRQ  & 767 & 4.5 & $5.47 \pm 0.15$ & $0.81\pm0.02$ & Uniform & 0.51 & 14 & 0.6--2.3 & -0.2--3.0 \\ \hline \hline
\end{tabular}
\end{adjustbox}
\end{center}
\flushleft
\footnotesize{$^{a}$Estimated in the polarization angle range [-90$^{\circ}$, 90$^{\circ}$) to avoid the alignment degeneracy.

$^{b}$Calculated before host galaxy or {BLR+AD} correction for host-galaxy dominated blazars and FSRQs, respectively.

$^{c}$Not enough number of measurements to reliably test the Von Mises distribution.

$^{d}$This source shows a double peak in the orientation in the EVPA. Due to its position, far from the edges of the distribution, the orientation of this source was evaluated with a double Gaussian function instead of a Von Mises distribution.}
\end{table*}

\subsection{Orientation of the EVPA}
\begin{figure*}
\centering
    \includegraphics[width=0.94\textwidth]{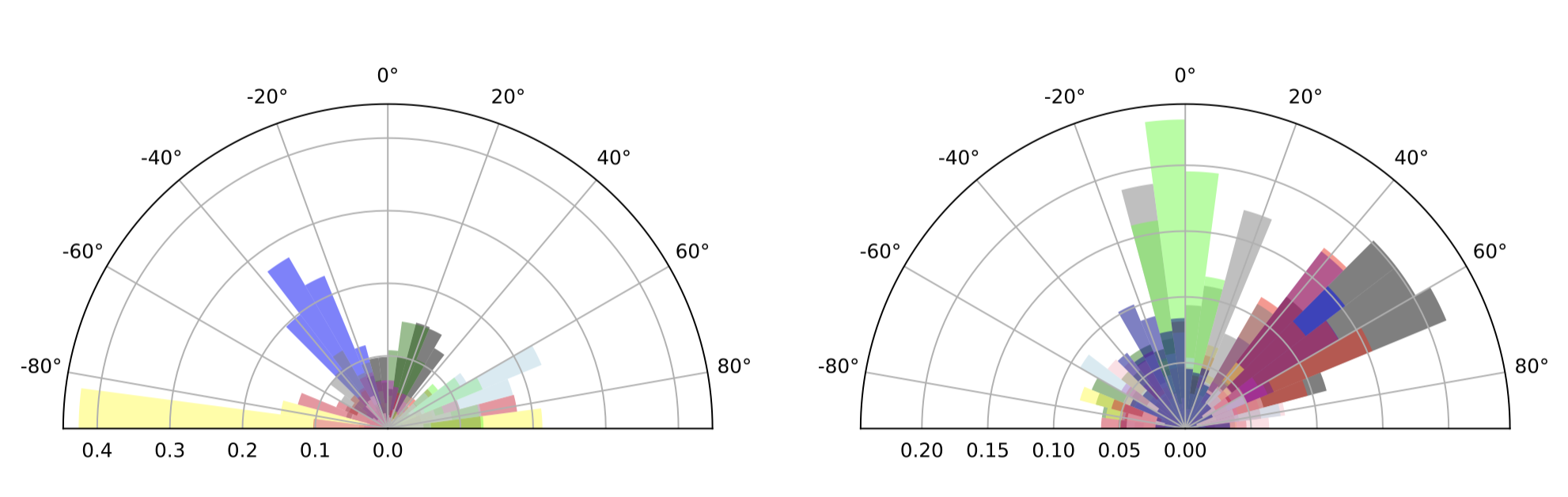}
    \caption{EVPA orientation histograms for the analysed blazar sample. Each colour represents the EVPA of one source. \textit{Left:} BL Lac objects. The EVPA tends to show a preferential orientation for each source. \textit{Right:} FSRQs. A more uniform or random distribution can be observed except for 3 objects at approximately 0$^{\circ}$ and 50$^{\circ}$.}
    \label{EVPA_orientations_polar}
\end{figure*}

We have also studied the distribution of the EVPA. {We have evaluated whether the sources analyzed here show a preferred or a random and stochastic orientation of the EVPA.} First, we note that the orientation of the polarization angle suffers from a $\pm$180$^{\circ}$ ambiguity introduced by the alignment degeneracy, causing that, for instance, $90^{\circ}=270^{\circ}$. We have taken into account this degeneracy by evaluating the distribution of the polarization angle in the range [-90$^{\circ}$, 90$^{\circ}$). The evaluation of these distributions is included in Table \ref{tab:polarization_results}. We compare a {Von Mises} distribution with a preferential orientation {in the considered angle interval} and a uniform distribution with no value of the polarization angle favoured. For the BL Lac subsample, 9 of the 11 sources present a dominant or preferred orientation of the polarization angle (see left panel of Figure \ref{EVPA_orientations_polar}). Contrary to this general behaviour we find AO~0235+164 and S5~0716+714. The former is know for showing mixed properties between the BL Lac and FSRQ classes \citep{raiteri2006,raiteri2014,otero-santos2022}. The latter on the other hand presents a low synchrotron peak in comparison with the bulk of the BL Lac population, as reported in Table~\ref{tab:blazar_sample}. FSRQs mostly present a random distribution of the polarization angle, as shown in the right panel of Figure \ref{EVPA_orientations_polar}. This is in line with the results presented by \cite{angelakis2016}, where a dependence between the frequency of the synchrotron peak and a preferential direction in the distribution of the EVPA is claimed. Therefore, FSRQs and BL Lacs with a low value of $\nu_{sync}$ tend to show a uniformly distributed polarization angle, whereas BL Lacs with a high $\nu_{sync}$ have a favoured orientation of the polarization angle. Nevertheless, four FSRQs of our sample still show a preferential orientation, i.e. PKS~1222+216, 3C~273, 3C~279 and 3C~345. We also note that, for three sources of our blazar sample, PKS~0735+178, H~1219+305 and H~1426+428, we have not tested their distribution due to the {low} number of observations.

\subsection{EVPA rotations}
The smooth EVPA rotation search performed according to the criteria established in Section \ref{sec:smooth_rotations} resulted in a total of {128} angle swings detected in our blazar sample. The results of this analysis are gathered in Table \ref{tab:rotations_results}. A total of {52} of the rotations were detected within the BL Lac subsample, while the {76} remaining swings correspond to the FSRQ population. We have not detected any rotation in the three blazars dominated by the stellar emission. All the smooth rotations are represented in Figures~\ref{h1426lc} to \ref{3c454lc} in the online version.

\begin{table}

\caption{Results of the EVPA rotation analysis on the different blazar types. (1) Source type. (2) Number of detected rotations. (3) Mean EVPA variation amplitude. (4) Mean rotation duration. (5) Mean EVPA variation rate. (6) Frequency of the rotations. {The uncertainties are estimated as the standard deviation of the distributions.}}
\label{tab:rotations_results}
\begin{adjustbox}{max width=\columnwidth}
\hspace{-0.4cm}
\begin{tabular}{cccccc}
\hline
(1) & (2) & (3) & (4) & (5) & (6) \\ 
\multirow{2}{*}{Type} & \multirow{2}{*}{N$_{rot}$} & $\langle \Delta \theta \rangle$  & $\langle T_{rot} \rangle$  & $\langle |\Delta \theta |/ \Delta t_{i} \rangle$  & $f_{rot}$  \\
 & & [$^{\circ}$] & [days] & [$^{\circ}$/days] & [days$^{-1}$] \\ \hline
BL Lacs & 52 & $144.7 \pm 9.1$ & $20.4 \pm 2.4$ & $19.5 \pm 3.1$ & 0.010 \\ \hline
FSRQs & 76 & $138.3 \pm 5.4$ & $21.1 \pm 1.5$ & $14.6 \pm 1.9$ & 0.017 \\ \hline
All sample & 128 & $140.9 \pm 4.9$ & $20.8 \pm 1.3$ & $16.6 \pm 1.7$ & 0.013 \\ \hline
\end{tabular}%
\end{adjustbox}
\end{table}

\begin{figure}
	\centering
	\includegraphics[width=\columnwidth]{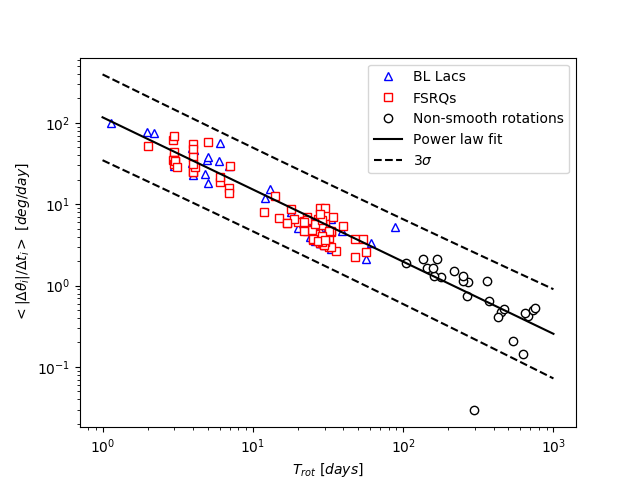}
	\caption{Rotation rate vs. duration of the observed rotations. Blue triangles and red squares correspond to BL Lac objects and FSRQs, respectively. Black circles correspond to the non-smooth slow rotations measured for all the blazar population. {The black solid line corresponds to a rotation of $|\Delta \theta|=117.3^{\circ}$.}}
	\label{fig:rate_vs_T_rotations}
\end{figure}

We have characterised each rotation by estimating their duration $T_{rot}$, amplitude of the EVPA change $\Delta \theta$ and variation rate, estimated as the mean rate of consecutive measurements in the rotation, $\langle |\Delta \theta_{i}| / \Delta t_{i} \rangle$. No significant differences are observed in the angle change measured for BL Lac objects and FSRQs, with both values compatible within errors. However, a small difference in the duration was found, with BL Lacs displaying slightly faster rotations than FSRQs. This leads to a faster variation rate measured for the former, $19.5 \pm 3.1$~$^{\circ}$/day, whereas the latter show a rate of $14.6 \pm 1.9$~$^{\circ}$/day. These rotations present a rather fast development, with the shortest one showing a duration of 1.1~days, while the slowest spans for 194.7~days approximately. By definition, the shortest possible swing corresponds to 90$^{\circ}$. Moreover, the largest EVPA change measured was 468.0$^{\circ}$. The relation between the rotation rate and the duration of the angle rotation is also represented in Figure \ref{fig:rate_vs_T_rotations}, {showing that rotations developing in shorter time scales present higher rate values, with a similar angle variation range as those occurring in longer time scales}. {This correlation between these quantities} is also observed in Figure 7 from \cite{blinov2016a} for the blazars monitored by the RoboPol programme. {The relation between the rate and the duration of the rotations} follows a power law function $\langle |\Delta \theta_{i}| / \Delta t_{i} \rangle = A \cdot T_{rot}^{-\alpha}$~$^{\circ}$/day, with $A=117.3^{\circ} \pm 40.8^{\circ}$ and $\alpha=0.89 \pm 0.01$ estimated from all the {smooth} rotations detected here. For the results presented here, we observe that all the {smooth} rotations are within the 3$\sigma$ limit of this power law fit. Visual inspection of this figure also reveals no significant differences between BL Lacs and FSRQs. Moreover, we have also evaluated a possible correlation of the rotation duration and variation amplitude, with no apparent relation between them. This was also reported by \cite{blinov2016a}. {Thus, a lack of correlation between the variation amplitude of the angle and the duration of the rotation is translated into a relation between the rate and the duration $\langle |\Delta \theta |/ \Delta t_{i} \rangle \propto T_{rot}^{-1}$. Therefore, the relation found between the rate and the duration with the power law function reported above are consistent with the expected behaviour.}

Given the total number of rotations detected, we can make an estimation for the frequency of the blazar sample of displaying an EVPA swing considering the procedure from \cite{blinov2016b}. These authors estimate the frequency as $f_{rot}=N_{rot}/N_{obs}$, where $N_{rot}$ and $N_{obs}$ are the number of rotations displayed and the number of days during which the sources were observed, respectively. This estimation was made for the full blazar sample, and for the BL Lac and FSRQ populations individually. The results are presented in Table~\ref{tab:rotations_results}. A frequency $f_{rot}=0.013$~days$^{-1}$ was estimated for the complete blazar population studied here. In comparison to the frequency estimated by the RoboPol programme {based on 24 sources showing EVPA rotations in the observations performed between 2013 and 2015} \citep{blinov2016b}, {the value estimated here} is roughly an order of magnitude larger. We note that the monitoring programme carried out by the Steward Observatory has a {faster mean observing cadence for several sources (shown in Table \ref{tab:polarization_results}) w.r.t. that from RoboPol during the first three seasons ($\sim$1~week), which could be leading to this difference. In fact, as pointed out by \cite{kiehlmann2021}, the frequency of observations plays a crucial role in the observations of EVPA rotations. These authors measured a frequency of EVPA rotations a factor 3 higher from season 4 of RoboPol (1-day cadence) than that estimated from the first three seasons.} 
Table~\ref{tab:rotations_results} also presents the frequency estimated for BL Lacs and FSRQs, with the latter showing a higher frequency than the former by a factor~$\sim$1.5.

Table \ref{tab:polarization_results} shows that eight blazars of the sample do not present any rotation during the monitored period. We have compared the characteristics of these eight non-rotating sources with those from the ones displaying EVPA swings. This comparison is presented in Table \ref{tab:rotators_nonrotators_differences}. The rotating population presents a higher polarization degree than those sources with no rotations in the evolution of their polarization angle, with mean values $P_{mean}=(8.32 \pm 0.97)$\% and $P_{mean}=(4.70 \pm 0.37)$\%, respectively. Moreover, they also show a higher maximum change of the EVPA, as well as a higher mean variation rate than the non-rotators by a factor $\sim$5.

\begin{table}
\centering
\caption[Characteristics of the rotator and non-rotator populations.]{Characteristics of the rotator and non-rotator populations. (1) Source type. (2) Mean polarization degree. (3) Maximum EVPA change. (4) Mean EVPA variation rate. {The uncertainties are estimated as the standard deviation of the distributions.}}
\label{tab:rotators_nonrotators_differences}
\begin{tabular}{cccc}
\hline
(1) & (2) & (3) & (4) \\ 
\multirow{2}{*}{Type} & $P_{mean}$  & $\Delta \theta_{max}$  & $\langle |\Delta \theta_{i}| / \Delta t_{i}\rangle$  \\ 
 & [\%] & [$^{\circ}$] & [$^{\circ}$/days] \\ \hline
Rotators & $8.32 \pm 0.97$ & $441.2 \pm 21.4$ & $16.6 \pm 2.1$ \\ \hline
Non-rotators & $4.70 \pm 0.37$ & $67.0 \pm 4.2$ & $3.6 \pm 0.9$ \\ \hline
\end{tabular}%
\end{table}

In addition, we have also investigated the differences in the behaviour of rotating blazars between rotating and non-rotating periods. {This comparison has been done for 103 rotations out of the 128 detected. We have discarded those cases in which the optical spectra from the Steward Observatory database are not flux calibrated.} We observe that for 12 of the 18 sources studied here presenting EVPA rotations, the mean polarization degree is lower during rotations that during their non-rotating periods. For two targets, this ratio has a value close to 1, while for the four sources remaining we measure a higher polarization degree during their rotations. Considering all the blazar sample, the mean polarization degree during rotations was estimated to be $P_{mean}=(6.50 \pm 0.68)$\%, lower than that during non-rotating observations with $P_{mean}=(8.44 \pm 1.02)$\%. This is in line with the results reported by \cite{blinov2016a} for 18 blazars studied in the RoboPol programme.

We have also evaluated the connections of smooth rotations with optical flares or enhanced flux states. We do not observe any systematic or general relation between flares and EVPA rotations for the bulk of the population, with several sources presenting polarization angle swings during both low and high emission states (for instance Ton~599). Therefore, no clear association can be made between these two types of events. However, we do observe that some sources show a clear increase of their mean optical flux during rotations. For instance, for 3C~345 we measured an average optical flux $F_{\lambda}^{mean}(\text{no-rotation})= (4.95 \pm 0.25) \times 10^{-16}$~erg~cm$^{-2}$s$^{-1}$ during non-rotating states, in comparison with $F_{\lambda}^{mean}(\text{rotation})=(1.11 \pm 0.10) \times 10^{-15}$~erg~cm$^{-2}$s$^{-1}$ during its rotation, roughly a factor 2 higher than the former. Other blazars showing this same behaviour are e.g. CTA~102 and OJ~248. In contrast to these sources, we also observe some of the targets of the sample with the opposite behaviour, i.e. a higher flux during non-rotations than during EVPA swings. Some examples of this are 3C~279 and AO~0235+164. For instance, the former has an optical flux $F_{\lambda}^{mean}(\text{rotation})=(1.80 \pm 0.18) \times 10^{-15}$~erg~cm$^{-2}$s$^{-1}$ and $F_{\lambda}^{mean}(\text{no-rotation})=(2.89 \pm 0.10) \times 10^{-15}$~erg~cm$^{-2}$s$^{-1}$ during rotations and non-rotations, respectively. We have also estimated the rate between the flux during rotating and non-rotating periods for the bulk of the blazar population included in this work. This rate yields a value of $F_{\lambda}^{mean}(\text{rotation})/F_{\lambda}^{mean}(\text{no-rotation})=1.24 \pm 0.12$, slightly higher for the former state, despite this behaviour not being the general rule for all the sources. This rate was also compared with that estimated by \cite{blinov2016a} for a different blazar population, with a value of 1.12, similar to the one estimated here.

Finally, as introduced in Section \ref{sec:nonsmooth_rotations}, we have also considered those rotations developed in longer time scales than the aforementioned smooth EVPA swings, showing fluctuations and fast variability in the measured EVPA during the rotation. A total of {24} non-smooth rotations in the {9 of the} 26 blazars included in this study. Following the approach used for the smooth angle swings, we have characterised each of these rotations. The average duration of these events, $\langle T_{rot} \rangle = 361 \pm 41$~days, is significantly {longer} than that {of smooth rotations as} in Table \ref{tab:rotations_results}. The fastest non-smooth rotations display a duration comparable with the slowest smooth EVPA swings, while the rest are clearly slower, as represented in Figure \ref{fig:rate_vs_T_rotations}. The mean polarization angle variation during these events is also higher than for the smooth rotations, with $\langle \Delta \theta \rangle = 253.1^{\circ} \pm 19.1^{\circ}$, in comparison with $\langle \Delta \theta \rangle = 140.9^{\circ} \pm 4.9^{\circ}$ reported in Table \ref{tab:rotations_results}. As a consequence of this difference, the measured rate is lower for the non-smooth rotations than for the smooth ones, with $\langle |\Delta \theta |/ \Delta t_{i} \rangle=1.0 \pm 0.1$~$^{\circ}$/day. It is also clear that these slow and non-smooth rotations follow the same relation between the rate and the duration of the rotation estimated for the smooth swings, as represented in Figure \ref{fig:rate_vs_T_rotations}. {The values of power law fit after including also the non-smooth rotations were found to be $A=117.2^{\circ} \pm 33.4^{\circ}$ and $\alpha=0.89 \pm 0.01$, consistent with those derived when considering only smooth rotations. As for the smooth rotations, we observe in Figure \ref{fig:rate_vs_T_rotations} that all the non-smooth rotations are contained in the 3$\sigma$ confidence limit of the fit with the exception of a non-smooth rotation displayed by 3C~66A.}

\subsection{Connection with the NMF analysis from OS22}\label{sec:comments}
Taking advantage of the spectral decomposition performed in OS22 using the NMF algorithm, we have investigated interesting behaviours and features displayed by the sources included here with the components derived from this analysis. The NMF study allowed us to decompose the total optical spectra in a {small} number of components that we could easily associate with the different parts of the blazar contributing to the optical emission. Galaxy-dominated blazars were reconstructed with a stellar template accounting for the emission of the host galaxy and a power law that modeled the synchrotron emission of the jet. {For BL Lac objects we used} two to four power laws to account for the emission of the jet and its variability. Finally, FSRQs needed an extra component corresponding to a quasi-stellar template to model the emission of the BLR. For this latter blazar type, we sometimes identified a steep and almost constant power law associated with a bright accretion disc. 

In this work we have investigated the connection of these components with the features and characteristics of the polarization and its variability. Making use of the polarized optical spectra of each blazar provided by the Steward Observatory, we have calculated the polarized spectral index assuming a power law shape as $F_{\lambda}^{pol} \propto \lambda^{-\alpha}$. Moreover, following the methodology from OS22, we have estimated the minimum number of components needed to model the variability of the polarized spectra using the residual sum of squares method. We have estimated a total of two components to account for the observed variability {in all three blazar types}. This result is expected as the jet is the only part that contributes significantly to the observed polarized emission. This can be compared with the same estimations from OS22 for the total spectra, where BL Lacs also yielded a total of two components, as their {total optical} emission is coming mainly from the relativistic jet. Contrary, FSRQs needed three or four components according to this approach owing to their more complex morphology, {with a bright BLR and accretion disc, contributing only to the total optical emission}. 

\subsubsection{Galaxy-dominated blazars}\label{sec:4.5.1}
Galaxy-dominated blazars present a very bright unpolarized stellar emission, with a very low variability in the total flux OS22. The high dominance of this contribution masks the spectral variability of the total spectra, reflected in the short variability range of the total spectral index (for instance, $\alpha_{tot} \sim 1.35-1.45$ for Mkn~501). On the other hand, the spectral variability of the polarized spectra is much higher, since the only contribution to this emission is the relativistic jet. The polarized spectral index presents a much higher variability range, with indices between $\alpha_{pol} \sim 0.5-2.5$. The difference between $\alpha_{tot}$ and $\alpha_{pol}$ for these sources confirms that the bright stellar emission is masking the real variability of the optical {jet} emission. Moreover, this stellar emission is also having a depolarizing effect in the measured polarization degree, as mentioned in Section \ref{sec:depolarization}.

\subsubsection{BL Lac objects}
BL Lac objects display a similar variability range for the total and polarized spectral indices. The slopes derived from the polarized spectra are also {comparable to} those derived from the NMF analysis, and reported in Table 2 from OS22. However, no strong correlation is observed between the total and polarized spectral indices, with typical values of the linear correlation coefficient $r\sim0.3-0.4$. Nevertheless, when the EVPA rotations are detected simultaneously with an optical flare ({<10\%} of the rotations), we typically
observe the same spectral behaviour of the total and polarized spectra. The spectral indices $\alpha_{tot}$ and $\alpha_{pol}$ during these events show rather similar values, meaning that the radiation producing these might have a common origin. In addition to this general behaviour, we also observe interesting features in some of the BL Lac objects studied here. Some comments on each BL Lac objects are included in Appendix~\ref{appendixA2}.

\subsubsection{FSRQs}
Following the behaviour of BL Lac objects, FSRQs also show a similar spectral variability in the total and polarized spectra. The correlation between both indices is absent. Moreover, roughly 30\% of the EVPA rotations identified in these objects were found to be connected with optical flares. During these events, similar total and polarized spectral indices are observed, often coincident with a dominant NMF component. Moreover, we do also observe peculiar behaviours for some FSRQs. In Appendix \ref{appendixA3} we include some comments on the FSRQs of the sample.

\section{Discussion}\label{sec:discussion}

Several models have been developed in recent years to explain the behaviour of the polarized emission observed from blazars. The results and differences between FSRQs/LSPs and ISPs/HSPs {found here} are consistent with those reported by \cite{angelakis2016} and \cite{hodge2018}. 
This behaviour can be explained through the shock-in-jet scenario proposed by \cite{angelakis2016}.
This framework is based on a magnetic field with a helical structure plus a turbulent component. In such scenario, a relativistic shock propagating in the jet accelerates particles through diffusive shock acceleration \citep[see Figure~16 from][]{angelakis2016}. The particles cool down as they are advected away from the shock due to the synchrotron and inverse Compton radiation. The high-energy particles that are responsible for the emission at frequencies around or above the synchrotron peak are located downstream in a small volume where the magnetic field has a highly ordered component plus a strong turbulent and variable contribution generated by the shock. On the other hand, the emission at frequencies below the synchrotron peak comes from particles contained in a much larger volume dominated by the ordered magnetic field. Therefore, a higher and more variable polarization is expected for the former region, while the second is expected to be less variable. This model has explained the behaviour observed for the blazar sample monitored by RoboPol. Due to the location of the optical emission w.r.t. the synchrotron peak for FSRQs/LSPs and ISPs/HSPs, this model explains the higher polarization and variability for the former in comparison to the latter. In addition, it can naturally explain the different EVPA distributions for each type. The optical emission from FSRQs is produced in the region with the strong turbulent component. Therefore, a random orientation is expected. Contrary to this, the optical emission of BL Lacs comes mostly from the region with a highly ordered, stable magnetic field, leading to a more constant orientation of the EVPA. Therefore, this model is also able to explain the results observed here. This is in line with one of the possible explanations proposed by \cite{smith1996} for the wavelength dependence reported between different blazar types, with a more ordered and aligned magnetic field component radiating mostly in bluer wavelengths. In fact, \cite{smith1996} associates this dependence to reasons intrinsic to the emitting region and rules out that it is introduced by external causes (e.g. accretion disc).

Concerning the EVPA rotation study carried out here, the results are compatible with those reported by \cite{blinov2016b}, where a higher frequency of rotations is measured for FSRQs than for BL Lacs, as reported in Table \ref{tab:polarization_results}. It is also important to note that among the BL Lacs displaying rotations, almost all correspond to the LSP and ISP subtypes. We also observe that the results found here are compatible with the distinction of rotations proposed by \cite{blinov2015} and \cite{kiehlmann2016}, with deterministic smooth rotations occurring in the dominant large-scale magnetic field; and stochastic rotations taking place downstream in the turbulent region. While the former can take place in both blazar types (however are expected to be more important in HBLs), the latter would be observed more often for FSRQs and LSPs. Moreover, owing to the higher contribution of the turbulent component for FSRQs, non-smooth rotations can also be related to this region, where short time-scale fluctuations are more likely to appear in the measured EVPA during the rotation. This is compatible with the fact that most of the non-smooth rotations observed here correspond to FSRQs. {The shock-in-jet model from \cite{angelakis2016}} can also explain the higher flux, variability and polarization degree measured for blazars displaying EVPA rotations in comparison to those where no rotations were observed. Indeed, rotations are more frequently observed in FSRQs and LBLs, which tend to be more variable and luminous than HSPs (see Figure~\ref{fig:intrinsic_modulation_vs_fvar}). The higher variability of the polarization for these sources w.r.t. HSPs is also explained through the stronger contribution of the turbulent component under this model, also explaining why FSRQs reach higher values of the polarization degree. Moreover, \cite{kielhmann2017} finds random walk models to be incompatible with the origin and development of EVPA swings, disfavouring turbulent models like the one proposed by \cite{marscher2014}.

Focusing on those blazars displaying EVPA rotations, we also observe differences between the periods in which a rotation is observed from those with no EVPA swings. In {agreement} with the results reported by \cite{blinov2016a,kielhmann2017}, the polarization degree was found to decrease during rotating periods. \cite{zhang2014,zhang2016} claim that this is an expected feature in a jet with a shock passing through the emitting region. However, random walk models are also able to reproduce such behaviour, as well as a flux increase during the rotations \citep[see for instance][]{blinov2016a}. Here we observe a marginal flux increase during this periods by a factor 1.24. However, this behaviour is not systematically observed for our blazar sample. Moreover, both of these scenarios are able to explain the absence of correlation between the amplitude of the polarization angle variability and the duration of the rotations. {\cite{kielhmann2017} also arise the possibility of the differences between rotating and non-rotating periods to be produced in random walk models.}

Rotations have been associated in the past with the development of $\gamma$-ray flares, with significant {evidence for} this connection reported by \cite{blinov2015,blinov2018}. We have also evaluated the possible relation between optical flares and EVPA rotations for our sample. As reported in Section \ref{sec:results}, we measure a hint of flux increase during rotating periods by a factor 1.24. Nevertheless, this brightening does not occur in all the blazars of our sample. In fact, a variety of different behaviours is observed, with some sources showing a clear flux increase (e.g. 3C~345), some with no significant change of their optical emission (BL~Lacertae), and some showing a lower flux (for instance 3C~279). We do not find any significant correlation between optical flares and the EVPA rotations measured here, in agreement with the results reported by \cite{blinov2016a} for the RoboPol blazar sample. The shock-in-jet model explains this as a random walk variability of the polarization if the rotations are being produced by the turbulent region \citep{blinov2016a}. Alternatively, if the shocks propagating along the jet are mildly relativistic, this model also predicts large variability of the polarization and the EVPA, leading to the observed rotations, with small flux variations \citep{zhang2016}. For those rotations that are observed correlated with an optical flare, as reported in Section \ref{sec:results}, the variability of the total and polarized spectral indices is approximately the same, meaning that the mechanisms leading to the flux and polarization variability may be related. In addition, we can also relate these events to the different components derived from the NMF reconstruction presented in OS22. We observe that BL Lac objects tend to show a higher correlation between the total and the polarized spectral indices than FSRQs. This can also be related to the fact that FSRQs show a higher dominance of the turbulent component and therefore, as aforementioned, their polarization may present a random walk variability. Contrarily for BL Lacs, dominated by the large-scale, stable helical magnetic field component, the variability mechanisms are more likely to be the same for the total and polarized emission.

Under the characteristics and behaviors detailed above, we find that the shock-in-jet model from \cite{angelakis2016} can provide a reasonable explanation for the reported results. However, it is important to note that this interpretation was made using the averaged properties extracted from all the blazar sample studied here. We also stress that more models have tried to reproduce the variability of the polarization and the EVPA. Some of them are models involving changes in the jet orientation \citep[for instance][]{marscher2008,lyutikov2017} or kink instabilities leading to a re-structuring of the magnetic field \citep[e.g.][]{nalewajko2017}. However, each model presents certain limitations when reproducing some of the observed features. For instance, the former example is expected to produce a determined direction of the rotation, something not observed here. For the second type of model highlighted above, these rotations would be limited to a maximum change of the EVPA of 180$^{\circ}$, always expected to happen in the same direction. Therefore, our observations disfavour these characteristics. Finally, models based on turbulent cells like the one proposed by \cite{marscher2014} have also been successful in the past for explaining the characteristics of the polarization variability. Nevertheless, large EVPA swings are rarely observed in such scenario, as well as not expected to appear correlated with MWL flares. We also note that models based on shocks propagating along helical jets similar to the one proposed by \cite{angelakis2016} have also been proposed for other sources such as S5~0716+714 and CTA~102 by \cite{larionov2013,larionov2016}, interpreting successfully the long-term behaviour of the polarization of several blazars.


\section{Conclusions}\label{sec:conclusions}
We have analysed 10 years of spectropolarimetric data for 26 $\gamma$-ray blazars monitored by the Steward Observatory, studying the properties of the polarization and its variability. Our results point towards a clear difference between FSRQs/LSPs and ISPs/HSPs. The former population presents a higher variability and reaches higher values of the polarization degree than the latter. Moreover, FSRQs tend to show a random distribution of the EVPA, in comparison to the preferential orientation displayed by BL Lacs. Concerning the relation between the flux and the polarization degree, BL Lacs do not show any correlated variability. On the other hand, FSRQs do appear to display a correlation between them. However, this correlation disappears after considering the depolarizing effect introduced by the~{BLR+AD}.

We have also performed a systematic search for EVPA rotations in the evolution of the polarization angle. This study also led to the observation of differences between FSRQs and BL Lacs, with more frequent rotations for the former. These rotations have not been statistically connected to optical flares. We have compared the characteristics of the blazars showing EVPA rotations with those with no polarization angle swings, observing a higher variability, flux and polarization for FSRQs. In addition, among the sources with EVPA rotations we have also detected differences between periods with ongoing rotations and periods were no rotations were taking place. The polarization degree measured during EVPA rotations was found to be lower than during non-rotating periods. Moreover, a marginal increase of the optical flux was also observed. However, this last effect does not appear systematically in all the blazars of the sample, as derived from the lack of correlations between rotations and flares.

Finally, we have evaluated the averaged characteristics of the polarization and its variability in the context of different models proposed in the literature. We conclude that the shock-in-jet model proposed by \cite{angelakis2016} can provide a plausible explanation for the observed average behaviour. This model assumes a large-scale helically ordered magnetic field, plus a turbulent magnetic field component. The former dominates the emission at frequencies below the synchrotron peak frequency, becoming more important in the optical emission for ISPs and HSPs; while the latter becomes stronger at frequencies around or above $\nu_{sync}$, dominating the optical emission in the case of FSRQs and LSPs. 

\section*{Data Availability}

All the data are publicly available at the Steward Observatory blazar monitoring programme webpage: \url{http://james.as.arizona.edu/~psmith/Fermi/}.


\section*{Acknowledgements}
JOS thanks the support from grant FPI-SO from the Spanish Ministry of Economy and Competitiveness (MINECO) (research project SEV-2015-0548-17-3 and predoctoral contract BES-2017-082171).
JOS, JAP and JBG acknowledges financial support from the Spanish Ministry of Science and Innovation (MICINN) through the Spanish State Research Agency, under Severo Ochoa Programme 2020-2023 (CEX2019-000920-S) and the project PID2019-107988GB-C22.
OGM acknowledges financial support from the project [IN105720] DGAPA PAPIIT from UNAM. 
Data from the Steward Observatory spectropolarimetric monitoring project were used. This programme is supported by Fermi Guest Investigator grants NNX08AW56G, NNX09AU10G, NNX12AO93G, and NNX15AU81G.
We thank the anonymous referee for his/her comments that helped to improve the manuscript.



\bibliographystyle{mnras}
\bibliography{biblio} 



\section*{Supplementary material}

{We include the polarization-flux diagrams for all the sources of the work, except those presented as an example in Section~\ref{sec:general_behaviour}. For galaxy-dominated blazars and FSRQs, we include the diagrams before and after correcting the depolarizing effect. We also include the long-term polarization light curves of the blazars studied here. \\}

\noindent{Figure S1\labtext{S1}{fig:polarization_galdom_online}}. Polarization-flux diagrams for galaxy-dominated blazars (same description as Figure~\ref{fig:polarization_galdom_mkn501}). 

\noindent{Figure S2\labtext{S2}{fig:polarization_bllacs_online}}. Polarization-flux diagrams for BL Lac objects. 

\noindent{Figure S3\labtext{S3}{fig:polarization_fsrqs_online}}. Polarization-flux diagrams for FSRQs (same description as Figure~\ref{fig:polarization_fsrq_3c454}). 

\noindent{Figure S4\labtext{S4}{h1426lc}}. Long-term polarization light curves of H~1426+428. 

\noindent{Figure S5\labtext{S5}{mrk501lc}}. Long-term polarization light curves of Mkn~501. 

\noindent{Figure S6\labtext{S6}{1es2344lc}}. Long-term polarization light curves of 1ES~2344+514. 

\noindent{Figure S7\labtext{S7}{3c66alc}}. Long-term polarization light curves of 3C~66A. 

\noindent{Figure S8\labtext{S8}{ao0235lc}}. Long-term polarization light curves of AO~0235+164. 

\noindent{Figure S9\labtext{S9}{s5_0716lc}}. Long-term polarization light curves of S5~0716+714. 

\noindent{Figure S10\labtext{S1}{pks0735lc}}. Long-term polarization light curves of PKS~0735+178. 

\noindent{Figure S1\labtext{S11}{oj287lc}}. Long-term polarization light curves of OJ~287. 

\noindent{Figure S12\labtext{S12}{mrk421lc}}. Long-term polarization light curves of Mkn~421. 

\noindent{Figure S13\labtext{13}{wcomlc}}. Long-term polarization light curves of W~Comae. 

\noindent{Figure S14\labtext{S14}{h1219lc}}. Long-term polarization light curves of H~1219+305. 

\noindent{Figure S15\labtext{S15}{1es1959lc}}. Long-term polarization light curves of 1ES~1959+650. 

\noindent{Figure S16\labtext{S16}{pks2155lc}}. Long-term polarization light curves of PKS~2155-304. 

\noindent{Figure S17\labtext{S17}{bllaclc}}. Long-term polarization light curves of BL~Lacertae. 

\noindent{Figure S18\labtext{S18}{pks0420lc}}. Long-term polarization light curves of PKS~0420-014. 

\noindent{Figure S19\labtext{S19}{pks0736lc}}. Long-term polarization light curves of PKS~0736+017. 

\noindent{Figure S20\labtext{S20}{oj248lc}}. Long-term polarization light curves of OJ~248. 

\noindent{Figure S21\labtext{S21}{ton599lc}}. Long-term polarization light curves of Ton~599. 

\noindent{Figure S22\labtext{S22}{pks1222lc}}. Long-term polarization light curves of PKS~1222+216. 

\noindent{Figure S23\labtext{S23}{3c273lc}}. Long-term polarization light curves of 3C~273. 

\noindent{Figure S24\labtext{S24}{3c279lc}}. Long-term polarization light curves of 3C~279. 

\noindent{Figure S25\labtext{S25}{pks1510lc}}. Long-term polarization light curves of PKS~1510-089. 

\noindent{Figure S26\labtext{S26}{b2_1633lc}}. Long-term polarization light curves of B2~1633+38. 

\noindent{Figure S27\labtext{S27}{3c345lc}}. Long-term polarization light curves of 3C~345. 

\noindent{Figure S28\labtext{S28}{cta102lc}}. Long-term polarization light curves of CTA~102. 

\noindent{Figure S29\labtext{S29}{3c454lc}}. Long-term polarization light curves of 3C~454.3.

\appendix

\section{Individual remarks}\label{appendix}
Here we include individual remarks for the sources considered in the polarization analysis carried out in this study.

\subsection{Galaxy-dominated Sources}\label{appendixA1}
\paragraph*{H 1426+428}
This is the blazar with the lowest observed polarization fraction of the three galaxy-dominated sources, with values <2\% prior to the host galaxy correction, and reaching fractions of $\sim$3.5\% after accounting for the depolarizing effect. Its flux is rather stable during its short monitored period (see Figure \ref{h1426lc}). No EVPA rotations where observed in the variability of the EVPA for this source. The total and polarized spectral indices adopt different values, with a higher variability of the latter ($\alpha_{pol}\sim -0.3-2.5$) in comparison to former ($\alpha_{tot}\sim 1.3-1.4$). This is due to the dominance of the host galaxy on the total spectra, as reported in Section \ref{sec:4.5.1}, masking its real variability.

\paragraph*{Mkn 501}
It is the only galaxy-dominated blazar showing EVPA swings. It displays a slow non-smooth rotation lasting $\sim$500~days at the beginning of the monitoring programme, with a swing of more than 100$^{\circ}$ (see Figure \ref{mrk501lc}). After this rotation, the EVPA stays rather stable -50$^{\circ}$. As for the other galaxy-dominated sources, the polarization degree of Mkn~501 is rather low in comparison with BL Lac objects and FSRQs, as well as its variability.

\paragraph*{1ES 2344+514}
As for the other two galaxy-dominated blazars, 1ES~2344+514 shows a low polarization fraction (<6\% before host galaxy correction) and a very stable EVPA. Its polarized flux shows one of the lowest variability among the blazar sample, ranging from $\sim$3~$\times~10^{-15}$~erg~cm$^{-2}$s$^{-1}$ to $\sim$4~$\times~10^{-15}$~erg~cm$^{-2}$s$^{-1}$ (see Figure \ref{1es2344lc}). The total spectral index also shows a very narrow variability range ($\alpha_{tot}\sim 0.8-1.0$), in contract with the highly variable polarized index ($\alpha_{pol}\sim 0.0-2.5$).

\subsection{BL Lac Objects}\label{appendixA2}
\paragraph*{3C 66A}
This object displays two non-smooth rotations around MJD~56600 and MJD 57050, approximately, one of them deviating from the power law fit describing the relation between the amplitude and the rate of the rotations detected here. 3C~66A displays preferred orientations of the EVPA. However, contrary to the rest of the BL Lacs with such behaviour, it presents a double peak in the EVPA distribution. During the period prior to the non-smooth rotations, the EVPA is {stable} at approximately 25$^{\circ}$. After these events, the peak is observed at -20$^{\circ}$. During these two events, $\alpha_{tot}$ and $\alpha_{pol}$ show a correlated variability, with a correlation coefficient $r=0.60$, with a p-value $=10^{-35}$. 

\paragraph*{AO 0235+164}
Roughly the same variability range is observed for $\alpha_{tot}$ and $\alpha_{pol}$. However, there is no correlation between both indices, following the FSRQ-like behaviour rather than the mild correlation shown by BL Lacs. This is consistent with the fact that this blazar has {shown} several times mixed properties between both blazar classes \citep{raiteri2007,raiteri2014}. The NMF reconstruction derives a component consistent with the presence of a BLR visible during low flux states (see OS22). Another indication of this comes from the uniform EVPA distribution displayed by this source, more typical of FSRQs.

\paragraph*{S5 0716+714}
This BL Lac object presents several EVPA rotations in its long-term evolution, as reported in Table \ref{tab:polarization_results} (see Figure~\ref{s5_0716lc}). As an example, we highlight one of these events, occurring on MJD~57000 simultaneous to optical flares. The polarized spectral index measured during this event ranges between $\alpha_{pol} \sim  0.9-1.2$. We observe that this variation coincides with that from the two NMF components dominating the emission during the flare, C1 and C2 with indices $\alpha_{1}=0.89$ and $\alpha_{2}=1.17$, respectively. This points towards a common origin of these components, the optical flare, and the observed polarization variability. The variability of $\alpha_{pol}$ for this source is consistent with that shown by both $\alpha_{tot}$ and the indices of the NMF components derived in OS22.

\paragraph*{PKS 0735+178}
This BL Lac is one of the least monitored during the 10-year period considered here. Its variability is also low, and the EVPA varies between 0$^{\circ}$ and 100$^{\circ}$ approximately, however without displaying any EVPA rotation (see Figure \ref{pks0735lc}).

\paragraph*{OJ 287}
The evolution of the EVPA for this object shows a slow decreasing trend, varying from roughly 180$^{\circ}$ to 120$^{\circ}$ in approximately 2000~days. After this period on MJD~57000, and anticipating a bright flare occurring on MJD~57300, a counter-clockwise non-smooth rotation takes place, leading to an orientation of the EVPA at $\sim$360$^{\circ}$ (see Figure \ref{oj287lc}). Then, simultaneously to the flare, a second clockwise non-smooth rotation happens, leading to a total EVPA swing of more than 400$^{\circ}$. During these events, the total and polarized indices vary correlated with values $\sim$1.0, compatible with the slopes of the NMF components that dominate during the flare (see OS22). This BL Lac displays a very clear correlation between $\alpha_{tot}$ and $\alpha_{pol}$, with a linear correlation coefficient $r=0.70$ (p-value $=10^{-75}$). Both indices vary roughly in the same interval, coincident with that from the different NMF components ($\alpha \sim 0.5 - 1.2$). 

\paragraph*{Mkn 421}
Several rotations are observed in the evolution of the EVPA in Mkn~421.{One} of them happens simultaneously with an optical {flux increase} on MJD~55600, {while another one is observed right before the  highest emission detected for this source in this 10-year period on MJD~56400, approximately, as} represented in Figure~\ref{mrk421lc}. As reported in OS22, {during this event} the component C2 from the NMF reconstruction dominates the emission of this flare, with $\alpha_{2} = 2.15$. This index is compatible with those measured during the rotation, $\alpha_{pol}\sim1.8-2.2$, suggesting a common origin of the flare and the EVPA swing. 

\paragraph*{W Comae}
{Five} rotations are identified in the long-term evolution the EVPA of W~Comae (see Figure \ref{wcomlc}). The first {four} occur during a low emission state, when the contributions of the three NMF components derived in OS22 are comparable. The last one is detected during the flare that takes place on MJD~58250, approximately, and dominated by C2 ($\alpha_{2}=1.13$). The total index during this event changes from values around 0.5 before the flare to around 1.0, compatible with the index of C2. The polarized spectral index varies around 1.2, showing consistent values. 

\paragraph*{H 1219+305}
As for PKS~0735+178, the data sample of this BL Lac is rather low w.r.t. the most monitored blazars. We observe a low polarization degree ($P < 7$\%) and a stable EVPA. No EVPA rotations are detected in the polarization angle variability of H~1219+305.

\paragraph*{1ES 1959+650}
This source is one of the least variable BL Lac objects. It shows a stable EVPA and a rather low polarization degree ($\sim$2.5-9\%), as shown in Figure \ref{1es1959lc}. No EVPA rotations are observed for this source. This is consistent with the fact that some hints of the host galaxy were observed in OS22. The polarized spectral index shows values between 1.0 and 2.0, approximately, while the total spectral index varies between 0.6 and 1.5.

\paragraph*{PKS 2155-304}
The EVPA is mainly oriented in the angle range between 60$^{\circ}$ and 90$^{\circ}$, showing only a few rotations during the 10-year period. Despite the variability displayed by this BL Lac object, no apparent connections between the rotations, the polarization degree and the flux variability are observed. These rotations are represented in Figure~\ref{pks2155lc}. The polarized spectral index varies within a range $\alpha_{pol} \sim 1.0-2.5$, similar to the total spectral index, $\alpha_{tot} \sim 1.0-2.1$.

\paragraph*{BL Lacertae}
A clear preferred EVPA orientation of 13$^{\circ}$ is observed for this source. Moreover, a large number of rotations are observed for BL~Lacertae, being the object of the sample with the highest number of rotations identified (see Figure \ref{bllaclc}), as reported in Table~\ref{tab:polarization_results}. However, as for PKS~2155-304, no evident relation with the flux or the NMF components is observed, with several rotations appearing during both high and low emission states. Both the total and spectral indices adopt values between $0.0$ and $1.0$, approximately. However, no strong correlation is observed in their variability. 

\subsection{FSRQs}\label{appendixA3}
\paragraph*{PKS 0420-014}
Despite the short time coverage of this FSRQ, a clear smooth rotation starting on MJD~55150 appears (see Figure \ref{pks0420lc}), coincident with the highest emission state displayed by PKS~0420-014. During this event both spectral indices vary between $0.5$ and $1.0$. In addition, this rotation also occurs during an enhancement of the contribution coming from the component C2 of the NMF analysis, characterized by a consistent spectral index $\alpha_{2}=0.89$.

\paragraph*{PKS 0736+017}
Several rotations are identified in the evolution of the EVPA for this FSRQ. The first one occurs simultaneously to the highest emission detected for this object in the monitored period. During this event, the total and polarized indices display values around 1.0, compatible with those from the dominant C2 and C3 components from the NMF analysis of OS22 during the development of this flare ($\alpha_{2}=0.92$ and $\alpha_{3}=0.50$). The second rotation is identified along a second flare, less bright than the first one, but also dominated by C3 and showing compatible indices for the total and polarized spectra. The rest of the rotations are observed during low emission states. 

\paragraph*{OJ 248}
This FSRQ shows a clear correlation of the polarization degree with the total flux, with an almost unpolarized low state, and reaching values of $\sim$18\%  during the brightest state (higher than 20\% after correcting from the depolarizing effect of the {BLR+AD}). OJ~248 also presents two rotations, coincident with the two peaks of the bright flare on MJD~56500, as shown in Figure \ref{oj248lc}, during which the total and polarized spectral indices adopt approximate values between $\alpha_{tot} \sim 1.3-2.3$ and $\alpha_{pol} \sim 1.0-2.5$, respectively. These events are also coincident with the dominance of components C3 and C2 of the NMF reconstruction, respectively, with indices $\alpha_{2}=1.33$ and $\alpha_{3}=1.06$.

\paragraph*{Ton 599}
A clockwise rotation is observed correlated with the flare occurring on MJD~56070 approximately (see Figure \ref{ton599lc}), related with the component C3 of the reconstruction ($\alpha_{3}=1.74$). During this event, the spectral indices $\alpha_{tot}$ and $\alpha_{pol}$ vary between 1.3 and 2.0, approximately, consistent with the spectral index of C3. An increase of the polarization degree is also measured. In addition, several rotations are detected in the low activity state right before and after the bright flare observed on MJD~58100, as shown in Figure \ref{ton599lc}.

\paragraph*{PKS 1222+216}
Contrary to the bulk of the FSRQ population, this blazar shows a clear preferential orientation of the EVPA at 0$^{\circ}$. It also shows {four} rotations, {three of them} linked to an increase of the flux of the reconstructed components C2 and C4 of the NMF analysis from OS22, between MJD~55600 and MJD~56100 (see Figure \ref{pks1222lc}). This components have a spectral index $\alpha_{2}=1.70$ and $\alpha_{4}=0.52$. The polarized spectral index shows values of $\alpha_{pol} \sim 0.3-2.1$ in the period during which the rotations occur. However, the total spectra show somewhat higher indices ($\alpha_{tot}\sim 2.2-3.3$). This could be due to the presence of a bright accretion disc, as reported in OS22, that makes the total spectra steeper than the polarized spectra, where the only expected relevant contribution is the synchrotron emission of the jet.

\paragraph*{3C 273}
As can be seen from Figure \ref{3c273lc}, 3C~273 is the least polarized blazar of the sample, with polarization degrees <1.6\%. This is also one of the few FSRQs showing a stable EVPA, showing almost no variability in both the polarization degree and angle, as well as in its total optical emission, as reported in OS22. This peculiar behaviour for a FSRQ, typically more variable and more polarized than BL Lac objects, may be due to the fact that the optical emission of this source is dominated by its bright accretion disc, as reported by \cite{raiteri2014} and in OS22. Therefore, the very low contribution to the polarized emission may be due to a very faint synchrotron emission, as well as a possible contamination from the interstellar medium or the accretion disc itself. Given the vary low variability displayed by the EVPA, no rotations are observed for 3C~273.

\paragraph*{3C 279}
This FSRQ is the only source of this type showing a clear correlation between $\alpha_{tot}$ and $\alpha_{pol}$. The measured linear correlation coefficient for these indices is $r=0.62$ (p-value$=10^{-53}$). Moreover, it is also one of the few FSRQs with a preferred EVPA orientation, contrary to the commonly observed uniform distribution for these sources. The EVPA is oriented towards $\sim$50$^{\circ}$, as reported in Table~\ref{tab:polarization_results}. One of the rotations observed for 3C~279 appears coincident with the enhanced state measured at MJD~58150 (see Figure \ref{3c279lc}). The NMF reconstruction from OS22 reports a dominance of components C2 and C4, with $\alpha_{2}=1.20$ and $\alpha_{4}=1.05$. The polarized spectral index varies around $\alpha_{pol}\sim1.25$ during this rotation, similar to the components contributing the most to the optical emission, and suggesting a connection of this rotation and the flare. \cite{larionov2020} reports a predominance of a helical magnetic field, which could be in line with the preferred orientation shown by 3C~279.

\paragraph*{PKS 1510-089}
The highly variable polarized emission of PKS~1510-089 shows a large number of rotations. Some of these events are found connected to optical flares, mostly coming from the same behaviour in the NMF component C3 ($\alpha_{3}=2.52$). However, this is not a general behaviour for this FSRQ, as some rotations are also observed during low emission states. It also presents non-smooth rotations, following the same rate-duration relation represented in Figure~\ref{fig:rate_vs_T_rotations}.

\paragraph*{B2 1633+38}
Several smooth rotations are observed in the highly variable behaviour of the polarization angle displayed by B2~1633+38, as shown in Figure \ref{b2_1633lc}. However, no clear association with flaring states or NMF components is observed. Moreover, {four} slow and non-smooth rotations are also present in the evolution of the polarization of this object. The first one takes place on MJD~56000, approximately. {The first of the three remaining starts} on MJD~56700 and the next two occur consecutively, with a duration of $\sim$500-800~days and swings >300$^{\circ}$. As for the case of PKS~1222+216, this FSRQ also shows total spectral index values ($\alpha_{tot}\sim 1.3-3.0$) slightly higher than those derived from the polarized spectra ($\alpha_{pol} \sim 0.1-2.2$). Again, this could be due to the presence of a bright accretion disc affecting the total spectral index values, as reported in OS22.                

\paragraph*{3C 345}
It shows an orientation of the EVPA towards $\theta \sim -52^{\circ}$. However, the low amount of data may be introducing a bias in the determination of the polarization angle distribution. Moreover, it shows a fast, smooth rotation coincident with its brightest flare, at MJD~55100. This EVPA swing also corresponds with an enhancement of the component C2 of the NMF reconstruction, with a spectral index $\alpha_{2}=1.39$. The total and spectral indices adopt similar values between 1.4 and 2.0 approximately during this event, compatible with that from the NMF component C2 responsible for the flare.

\paragraph*{CTA 102}
This blazar shows the typical lack of correlation between the total and polarized spectral indices observed for FSRQs. During a bright flare corresponding to the brightest optical state observed in this monitoring, occurring on MJD~57750 approximately, {two} EVPA rotations {are} also observed (see Figure \ref{cta102lc}). As reported in OS22, this flare is described with the component C2, with $\alpha_{2}=1.23$. The measured polarized spectral index in this period is similar to that derived of C2, with $\alpha_{pol}\sim 1.5$. Another important feature of CTA~102 is a clear increase of the polarization degree during its EVPA rotations, contrary to the behaviour reported for the bulk of the population.

\paragraph*{3C 454.3}
This is one of the most variable sources of the sample, with several bright flares dominating its optical emission. The spectral indices $\alpha_{tot}$ and $\alpha_{pol}$ show no correlation in their variability. During two enhanced optical states occurring at MJD~55200 and MJD~56800, we observe {three} EVPA rotations that happen simultaneously. {One of them is coincident with a minor flare on MJD~55200, and the two remaining are coincident with the brightest flare on MJD~56850, approximately.} These events can be seen in Figure \ref{3c454lc}. The spectral variability of these events suggest, as already claimed for other sources, a common origin of the flux and polarization variability, with total and polarized spectral indices $\alpha_{3}=0.89$ and $\alpha_{pol}\sim 1.0$ for the first flare, and $\alpha_{2}=0.54$ and $\alpha_{pol}\sim 0.5$ for the second one, where $\alpha_{3}$ and $\alpha_{2}$ correspond to the indices of the NMF components C3 and C2, respectively. These components are the ones dominating the emission during each of these flares, as reported by OS22. 




\bsp	
\label{lastpage}
\end{document}